\documentclass[11pt,a4paper]{article}

\usepackage{amsmath}
\usepackage{amsthm}
\usepackage{amssymb}
\usepackage{amsfonts}
\usepackage{wasysym} 
\usepackage{cancel}
\usepackage{dsfont} 
\usepackage{mathrsfs}
\usepackage{mathtools}
\usepackage{graphics}
\usepackage[utf8]{inputenc}

\usepackage[a4paper,bottom=3.0cm,top=2.5cm,left=3.2cm,right=3.2cm]{geometry}

\newtheoremstyle{mystyle}
  {\topsep}{\topsep}{}{}{\bfseries}{ }{\newline}
  {\thmname{#1}\thmnumber{ #2}~\thmnote{(#3)}%
     \ifstrempty{#3}%
      {\addcontentsline{toc}{subsection}{#1~\thethm}}%
      {\addcontentsline{toc}{subsection}{#1~\thethm~(#3)}}%
  }

\newcommand{\fl}[1][3em]{\hspace{#1}&\hspace{-#1}} 

\newcommand{\nodate}{\date{\vspace{-1.0em}}}

\newcommand{\Ii}{\mathrm i} 
\newcommand{\Ee}{\mathrm e} 
\newcommand{\Alg}{\mathfrak A}

\newcommand{\term}[1]{{\it #1}}
\newcommand{\Reg}{\mathcal O}

\newcommand{\DoubleCone}{{\mathscr C}}

\newcommand{\RealNum}{\mathbb R}
\newcommand{\ComplexNum}{\mathbb C}

\newcommand{\NaturalNum}{\mathbb N}
\newcommand{\SchwartzSpace}{\mathscr S}
\newcommand{\DInt}[2][\,]{{\mathrm d}^{\hspace{-0.20ex}#1}\hspace{-0.25ex}#2}

\newcommand{\norm}[1]{\left \|#1 \right\|}
\newcommand{\normm}[1]{\|#1\|}
\newcommand{\abs}[1]{\left |#1 \right|}
\newcommand{\abss}[1]{|#1|}
\newcommand{\CStar}{\mathrm C^*}

\newcommand{\HilbertSpace}{\mathscr H}
\newcommand{\BoundedOps}{\mathrm B}

\newcommand{\LSpace}{L}

\newcommand{\Id}{\mathds 1}
\newcommand{\Ball}{{\mathrm B}}

\newcommand{\ContinuousFuncs}{\mathnormal C}

\newcommand{\IndicatorFunction}{\mathds 1}

\newcommand{\CharFct}{\IndicatorFunction}

\usepackage{xcolor}
\usepackage{hyperref}
\hypersetup{pdfborderstyle={/S/U/W 0},colorlinks=true,linkcolor=blue,
pdfnewwindow=true,linkcolor=blue!30!black!80,citecolor=blue!30!black!80,urlcolor=blue!30!black!80} 
\usepackage{cleveref}
\crefformat{footnote}{#2\footnotemark[#1]#3}

\usepackage{tikz}

\usepackage{etoolbox}

\usepackage{enumerate}

\newcommand{\uvec}[1]{\underline{#1}}
\renewcommand{\vec}[1]{{\bf #1}} 

\newcommand{\closure}[1]{\overline{#1}}

\DeclareMathOperator*{\support}{supp}

\DeclareMathOperator{\Dom}{D} 

\newcommand\numberthis{\addtocounter{equation}{1}\tag{\theequation}}

\newcommand{\standardThms}{\newcounter{thm}
\newtheorem{Prop}[thm]{Proposition}
\newtheorem{Lem}[thm]{Lemma}
\newtheorem{Cor}[thm]{Corollary}
\newtheorem{Thm}[thm]{Theorem}
\newtheorem{Def}[thm]{Definition}

\theoremstyle{remark}
\newtheorem{Rem}[thm]{Remark}
}

\usepackage[
  backend=bibtex,
  natbib=true,
  style=alphabetic,
  doi=true,
  isbn=false,
  url=false,
]{biblatex}

\renewbibmacro{in:}{}
\newcommand\itr[1]{\mbox{($#1$\hspace{1pt})}}

\newcommand{\FockSpace}{\mathscr F}
\newcommand{\SymmetricGroup}{\mathfrak S}

\newcommand{\s}{3}
\newcommand{\sOne}{4} 
\newcommand{\sPOne}{4} 
\newcommand{\sPOneH}{2} 
\newcommand{\sMOne}{2} 
\newcommand{\Uu}{\mathcal U}
\newcommand{\itref}[1]{\mbox{(\,{\!\it\ref{#1}\hspace{1pt}})}}
\def\T{\tau}

\newcommand{\be}{\beta}

\newcommand{\mco}{\Reg}

\newcommand{\Om}{\Omega}
\newcommand{\De}{\Delta}

\def\Bb{{B}} 
\def\Aa{{A}} 

\def\BB{\mathcal{B}} 
\def\AA{\mathcal{A}} 
\def\CAHR{C_{\text{\rm AHR}}}
\newcommand{\NAHR}[1]{\norm{#1}_{\text{\rm AHR}}}
\def\Tmax{\T_{\text{\rm max}}}
\def\Tmin{\T_{\text{\rm min}}}
\def\Pp{\boldsymbol{P}}
\def\gRS{\gamma_{\text{RS}}}

\def\Oc{,} 
\def\Oc{} 
\def\NOc{\,} 
\def\lss({(}
\def\rss){)}

\def\ff{{\bf f}}

\title{Strengthened Reeh-Schlieder Property 
  and Scattering in Quantum Field Theories without Mass Gaps}
  \def\affil{Zentrum Mathematik, Technische Universität München,\\[0.3em] D-85747
Garching, Germany}
\def\mail{duell@ma.tum.de}
\author{Maximilian Duell\footnote{E-mail: \mail}\\[1.3em]\affil}
\nodate
\standardThms

\def\tfamily{family}
\def\tfamilies{families}
\begin{document}
\maketitle

\begin{abstract}
We develop Haag-Ruelle scattering theory for Wigner particles
in local relativistic Quantum Field Theory without assuming mass gaps
or any other restrictions on the spectrum of the mass operator near the
particle masses. 
Our approach is based on the Reeh-Schlieder property
of the vacuum state. It is shown that a strengthened variant of this property,
concerning the relative approximation error for single-particle states, 
implies the existence of scattering states.

\end{abstract}

\section{Introduction}

The infrared problem in Quantum Electrodynamics
(QED) has attracted a lot of attention in the mathematical physics literature of
the last decade. Consistent
scattering theory has been developed for various physical processes involving
charged particles~(`electrons'), neutral massive particles~(`atoms') and
massless particles~(`photons'). Some of these results were obtained in non-relativistic
models of QED
\cite{CFP10, DyP13, MS14b}, others in the general setting of
algebraic QFT \cite{BR14, AD15,Dy05,Hdg13, DH14}.
In spite of all these efforts, even the seemingly simple case of scattering of
several atoms is still not fully under control.

This may be explained by the fact that atoms in QED constitute a prototypical example of
an {\em embedded}\/ particle. In other words, single-atom
states correspond to eigenvalues of the mass operator which are not isolated,
but embedded in a continuous mass spectrum, arising e.g.\ from states
consisting of multiple lighter particles (photons). 
For the construction of scattering states, such background particles need to be
separated from the desired single-atom states. 
In the framework of Haag-Ruelle theory, this separation could so far
only be achieved with the help of technical
assumptions\footnote{\label{foot:rel}See e.g.\ \cite{Hrb71, Dy05,
Hdg13,DH14}.} on the spectral measure of the mass operator near the particle
masses. Such spectral conditions were first proposed by Herbst
\cite{Hrb71} and we might consider them to be a remnant of the original
Haag-Ruelle mass-gap assumption \cite{Ha58, Ru62, Hep65}.  

As the
physical meaning of these assumptions has remained obscure, the existence of
scattering states of atoms still lacks a conceptually clear explanation. Aiming
at such an explanation, we develop Haag-Ruelle scattering theory for atoms,
relying on certain non-local correlations of the vacuum state. 
The required condition is only slightly stronger than the well-established
Reeh-Schlieder property and it permits to the best of our knowledge the first
proof of existence of scattering states of massive embedded Wigner particles
without a priori requiring a spectral condition of Herbst type.

The Reeh-Schlieder property states that the vacuum~\(\Omega\) is cyclic
for any algebra~\(\Alg(\Reg)\) of observables\footnote{In the case of QED these
  algebras should be generated by bounded functions of suitably smeared
  electromagnetic fields and the electric current, cf.\ \cite{Bu86}.}
localized in a bounded space-time region~\(\Reg\). That is, given any vector
\(\Psi\in \HilbertSpace\) (for example describing an atom essentially localized
far from the region \(\Reg\)), there exists a \tfamily\ of observables
\((\Aa_{\beta})_{\beta>0}\) from \(\Alg(\Reg)\) such that
\begin{equation}     
\lim_{\beta\to 0}\|\Aa_{\beta}\Omega-\Psi\|=0. \label{RS}
\end{equation}
While \(\norm{A_\beta\Omega}\) clearly remains bounded, we note that the
operator norms \(\norm{\Aa_{\beta}}\) may tend to infinity as
\(\beta\to 0\). 
%
%
As it will be important for our investigation to quantify this growth,
we will say that \(\Psi\) is a vector of finite \term{Reeh-Schlieder degree}
if there exists a \tfamily\ of operators~\((\Aa_{\beta})_{\beta>0}\) localized in
some fixed bounded space-time region \(\Reg\), such that
for some \(\gamma > 0\) we have
\[     
  \norm{\Aa_{\beta}}  \leq \beta^{-\gamma} 
  \quad\text{and}\quad
  \norm{\Aa_{\beta}\Omega - \Psi} \leq \beta.
  \label{RS1}\numberthis
\]
In this paper we will construct scattering states of configurations of atoms
whose single-particle states are generated by such \tfamilies\ with finite
Reeh-Schlieder degree \(\gamma\).
 Condition \eqref{RS1} is readily verified for free scalar fields\footnote{%
  A free scalar field \(\phi(f)\) is self-adjoint for real-valued \(f\) 
   and \(\phi(f)\Omega\), 
\(f\in\ContinuousFuncs^\infty_c(\RealNum^\sPOne)\),
  yield a dense subset of single-particle states.
  If \(\support f \subset \Reg\) we can simply set~\(A_\beta := \phi(f)
  \exp(-\beta \abss{\phi(f)}^{1/\gamma}) \in \Alg(\Reg)\) to obtain Reeh-Schlieder
  \tfamilies\ of arbitrarily small degrees \(\gamma > 0\).
  For further examples see \Cref{app:gff}.
  \label{foot:RSFF} 
  }, 
 but it seems that not much progress has been made in understanding such
 relations since the seminal work of Haag and Swieca~\cite{HS65}. 
 In theories where Herbst's spectral condition is satisfied, one can
 construct an operator family \((A_\beta)_{\beta > 0}\) satisfying a weakened
 variant \eqref{eq:condFlat}  of \eqref{RS1} (see concluding discussion),  but
 the status of \eqref{RS1} in interacting theories is currently not clear
 and constitutes a difficult technical problem  outside the scope of this work.


 Let us now describe in non-technical terms the relevance of \eqref{RS1} for 
 Haag-Ruelle scattering theory. Take a single-atom state \(\Psi\) of finite Reeh-Schlieder degree and let
\((\Aa_{\be})_{\be>0}\) be a corresponding \emph{Reeh-Schlieder \tfamily} from
formula~(\ref{RS1}). 
Since \((\Aa_{\be})_{\be>0}\) play a role of creation operators, it is technically
convenient to smear them with the Fourier transform of a function \(\hat \chi \in
\ContinuousFuncs^\infty_c(\RealNum^{\sOne} \setminus \bar V^-)\)
yielding a \tfamily\ of almost-local operators 
\begin{equation}
  \Bb_\beta 
  := \int \DInt[\sOne]x \; \chi(x) \Aa_\beta(x) 
  , \;\; (\beta > 0), \label{smearing-intro}
\end{equation}
where \(\Aa_{\beta}(x)\) denotes the translate of \(\Aa_{\beta}\) in
space-time by \(x\). 
Following  the standard prescription we pick  a regular positive-energy
solution \(f\) of the Klein-Gordon equation with the mass of the atom and set
\begin{align}
  \BB_\T &
  := \int \DInt[\s] x\, f(\T, \vec x) \Bb_{\beta(\T)}(\T,
  \vec x), \quad \text{with}\ \  \beta(\T) := \T^{-\mu}, \; \mu > 0 \;
  \text{fixed}. \label{AT}
\end{align}
We will call \(\BB_\T\) an (approximating) {\em creation operator} of \(\Psi\) since it has the
property
\begin{equation}
  \lim_{\T\to\infty}  \BB_\T\Omega 
  = (2\pi)^{\sPOneH} \hat \chi(H, \Pp) 
  \tilde f(\vec P) \left(\lim_{\T\to\infty} \Aa_{\be(\T)} \Om \; \right)
  =  (2\pi)^{\sPOneH} \hat \chi(H, \Pp)  \tilde f(\Pp)\Psi 
  \label{single-particle}
\end{equation}
(see \Cref{prop:elemProp}).
That is, it asymptotically creates \(\Psi\) from the vacuum up to an inessential
function of the energy-momentum operators \((H, \Pp)\) (which can be arranged
to be equal to one if \(\Psi\) has bounded energy).  Since we inserted a
Reeh-Schlieder \tfamily\ in (\ref{AT}), we obtain convergence in
(\ref{single-particle}) without the ergodic averaging used in earlier works
\cite{Dy05, Bu77}. We also note that (\ref{single-particle}) holds even for
\(\Psi\) of infinite Reeh-Schlieder degree. The need to assume finiteness
of the Reeh-Schlieder degree of \(\Psi\) arises only at the level of \(n\)-atom
scattering states, \(n\geq 2\) --- the case to which we now proceed.


Let \(\Psi_1, \Psi_2\) be two single-atom states with disjoint velocity supports
and finite Reeh-Schlieder degree. Let \(\BB_{1\Oc\T}, \BB_{2\Oc\T}\) be the
corresponding creation operators constructed as above. The
scattering state describing these two atoms is given by the limit as \(\T\to
\infty\) of the \tfamily\footnote{For clarity reasons we consider here only
outgoing states. The incoming case \(\T \rightarrow - \infty\)
is analogous.} 
\[
  \Psi_\T := \BB_{1\Oc\T}\BB_{2\Oc\T}\Om.
\]
The  conventional Cook-argument to establish convergence does not apply here
due to the additional \(\T\)-dependence via the Reeh-Schlieder \tfamily\ in
(\ref{AT}). Therefore, we base our proof on a discretized analog of Cook's
argument involving summability of the telescopic expansion
\[
  \norm{\Psi_{\T_N} - \Psi_{\T_0}}
  \leq \sum_{k=0}^{N-1} \norm{\Psi_{\T_{k+1}} - \Psi_{\T_k}}
  \label{eq:telescope}\numberthis
\]
in the limit \(N \rightarrow \infty\) (here \(\T_k := (1+\rho)^k \T_0, \, \T_0
> 0\), and \(\rho>0\) is sufficiently small). The first term in this sum has the
form
\begin{equation} \Psi_{\T_1}-\Psi_{\T_0}=\BB_{1\Oc\T_1}(  \BB_{2\Oc\T_1}-
  \BB_{2\Oc\T_0})\Om+ (\BB_{1\Oc\T_1} -  \BB_{1\Oc\T_0})\BB_{2\Oc\T_0}\Om.
\end{equation}
Exploiting locality and the fact that $|\T_1-\T_0|$ is small, we obtain that
\([(\BB_{1\Oc\T_1} -  \BB_{1\Oc\T_0}), \BB_{2\Oc\T_0}]\) is rapidly decreasing with \(\T_0\) and
thus it suffices to study the expressions
\begin{equation} 
  \BB_{1\Oc\T_1}( \BB_{2\Oc\T_1} - \BB_{2\Oc\T_0})\Om,
  \quad\quad  \BB_{2\Oc\T_0} (\BB_{1\Oc\T_1} - \BB_{1\Oc\T_0})\Om. \label{two-terms}
\end{equation}
Let us concentrate on the first term above: Thanks to the smearing
operation~(\ref{smearing-intro}) which restricts the energy-momentum transfers
of the creation operators, we can write
\begin{equation} \|\BB_{1\Oc\T_1}(  \BB_{2\Oc\T_1}-
  \BB_{2\Oc\T_0})\Om\| \leq \| \BB_{1\Oc\T_1}E(\De) \| \|(  \BB_{2\Oc\T_1}-
  \BB_{2\Oc\T_0})\Om\|, \label{submult}
\end{equation}
where \(E(\De)\) is a projection onto a compact subset \(\De\) of the energy-momentum
spectrum.  Now exploiting formula~(\ref{single-particle}) and results from
\cite{Bu90}, which give \(\|\BB_{1\Oc\T_1}E(\De) \|\leq
C\|\Aa_{1\Oc\be(\T_1)}\|\), we can estimate~(\ref{submult}) by
\begin{equation}
\|\Aa_{1\Oc\be(\T_1)}\|  \|\Aa_{2\Oc\be(\T_1)}\Om-\Aa_{2\Oc\be(\T_0)}\Om\|\leq
\|\Aa_{1\Oc\be(\T_1)}\| ( \|\Aa_{2\Oc\be(\T_1)}\Om-\Psi_2\| + \|\Aa_{2\Oc\be(\T_0)}\Om-\Psi_2\|)
\end{equation}
up to an overall constant, and the analysis of the second term in
(\ref{two-terms}) gives an analogous bound. By substituting such estimates into
(\ref{eq:telescope}), it is easy to obtain convergence of~\(\Psi_\T\), provided
\(\Psi_1, \Psi_2\) are of Reeh-Schlieder degree \(\gamma < 1\) (cf.\
relations~(\ref{RS1}), (\ref{AT})). A similar discussion of
\(n\)-atom scattering states could suggest that single-atom states of arbitrarily
small Reeh-Schlieder degree are needed. It turns out that this is not the case:
by careful geometrical analysis and application of corresponding novel
multi-operator clustering estimates (cf.\ Lemmas~\ref{lem:estNoneqB} and
\ref{lem:multiClust2}, respectively) we develop complete Haag-Ruelle scattering theory for
single-atom states of arbitrarily large Reeh-Schlieder degree. 
Although atoms are our prime example, the construction works equally well for
photons\footnote{In contrast to atoms, scattering theory of photons is well
understood since \cite{Bu77}.}, which demonstrates the robustness of our
approach. We hope that this investigation will pave the way to a definite
unifying solution of the problem of scattering of Wigner particles in algebraic
QFT.


This paper is structured as follows: 
In \Cref{sec:FWAss} we state the basic assumptions underlying this work and
introduce the {Reeh-Schlieder degree} of Hilbert-space vectors.
\Cref{sec:basic} gives an exposition of our variant of Haag-Ruelle
creation operators and establishes some of their basic properties.
\Cref{sec:commut} provides the fundamental technical tool of the discretized
Cook's method: we derive rapid norm decay of non-equal time commutators of
creation operators. In Sections~\ref{sec:cluM} and
\ref{sec:cluConseq} we establish clustering estimates and study their
consequences relevant for refined handling of the norm growth of the
creation operator approximants.
All these results are then combined in \Cref{sec:conv} to prove convergence of
scattering states and to establish their Fock structure in \Cref{sec:Fock}.



{
\paragraph{Acknowledgements}~\\
I am indebted to Klaus Fredenhagen for the suggestion to accelerate the
convergence in the single-particle problem via the Reeh-Schlieder
property.
Similarly I would like to thank Wojciech Dybalski for encouragement and numerous
insightful advice extended during the course of this work. 
Further I profited from helpful discussions with
Sabina Alazzawi, Detlev Buchholz, Maximilian Butz, Daniela Cadamuro, and Yoh
Tanimoto.
Financial support from the Emmy Noether Programme of the DFG (grant
DY107/2-1) is gratefully acknowledged.
}

\section{Framework and assumptions} \label{sec:FWAss}

As the basis for our considerations we take a  Haag-Kastler theory in the vacuum
representation, i.e.\ a net \(\Reg \longmapsto \Alg(\Reg) \subset
\BoundedOps(\HilbertSpace)\) of von Neumann algebras associated to bounded open
regions~\(\Reg \subset \RealNum^{\sPOne}\) in Minkowski space-time\footnote{We
take the space-time metric with signature $(+,-,-,-)$.}.
Space-time translations by vectors~\(x = (t, \vec x) \in
\RealNum^{\sPOne}\) are represented on the Hilbert space~\(\HilbertSpace\) by a
strongly-continuous group of unitary operators   \(U(t, \vec x)=\Ee^{\Ii t H -
\Ii \vec x \cdot \Pp}\), generated by the strongly-commuting family of 
the self-adjoint \emph{energy-momentum} operators~\((H, \Pp)\). Their joint
spectral measure is denoted by \(E(\Delta) :=
E_{(H,\Pp)}(\Delta)\) for any Borel set \(\Delta \subset \RealNum^{\sPOne}\).
The {\em vacuum}\/ is a normalized translation-invariant vector \(\Omega \in
\HilbertSpace\).
Finally, translations of operators \(A \in \BoundedOps(\HilbertSpace)\) are
induced by \(U\) according to \(A(x) := \alpha_{x}(A) := U(x) A U(x)^* \).
We will use the following version of the Haag-Kastler postulates,
\newcommand{\lt}[1]{\text{\bf #1}}%
{\allowdisplaybreaks%
\begin{align*}
  \lt{Isotony} \quad 
    & \Alg(\Reg_1) \subset \Alg(\Reg_2) \text{ for } \Reg_1 \subset \Reg_2 
  \tag{HK1} \label{eq:HK1}\\
  \lt{Locality} \quad 
  & \Alg(\Reg_1) \subset \Alg(\Reg_2)' 
\text{ for } \Reg_1 \subset \Reg_2' 
\tag{HK2} \label{eq:HK2}\\
  \lt{Covariance} \quad  
  & \alpha_x(\Alg(\Reg)) = \Alg(\Reg +x 
  )\tag{HK3} \label{eq:HK3}
\\
  \lt{
   Uniqueness of \(\boldsymbol \Omega\)} \quad  &
   E(\{0\}) \HilbertSpace = \ComplexNum \Omega \tag{HK4}
   \label{eq:HK4}\\
  \lt{Spectrum Condition} \quad 
  &
  \support E_{(H, \Pp)} \subset \bar V^+ \tag{HK5}
  \label{eq:sCond}\\
    \lt{Reeh-Schlieder Property} \quad &
   \closure{\Alg(\mco)\Omega} = \HilbertSpace\tag{HK6}
   \label{eq:RSProp}
\end{align*}
}%
for any non-empty open bounded regions \(\Reg, \Reg_1, \Reg_2 \subset \RealNum^{\sPOne}\)
and any \(x \in \RealNum^{\sPOne}\).
Here, \(\Alg(\Reg)'\) is the commutant of \(\Alg(\Reg)\) in
\(\BoundedOps(\HilbertSpace)\) and \(\Reg' := \{ y \in \RealNum^{\sPOne}: (y-x)^2 < 0
\;\forall x \in \Reg \}\) defines the causal complement of \(\Reg\).
Further, \(\bar V^\pm := \{ x \in \RealNum^{\sPOne} : x^2 \geq 0, \pm x^0 \geq 0 \}\) is the future or past 
light cone, respectively. For future reference we denote by \(\Alg\)
the \(\CStar\)-inductive limit  of the local net and by \(H_m:=\{ p \in \RealNum^{\sPOne} : p^0=\sqrt{\vec p^2+m^2} \}\) the mass hyperboloid of a particle of mass \(m\geq 0\).

Next, we define the \emph{Reeh-Schlieder degree} \(\gRS\geq 0\)
of a vector \(\Psi\in \HilbertSpace\) as the infimum over all \(\gamma\geq 0\)
for which there exists an open bounded region \(\Reg\) and a \tfamily\ of
observables \((\Aa_{\beta})_{\beta>0}\) from \(\Alg(\Reg)\) such that
for all sufficiently small \(\beta > 0\) we have
\begin{equation}     
  \qquad \norm{\Aa_{\beta}\Omega-\Psi} \leq \beta,
  \qquad \norm{\Aa_{\beta}} \leq \beta ^ {-\gamma}.
\label{RS1-section1}
\end{equation}
We will call \((\Aa_{\beta})_{\beta>0}\) a \emph{Reeh-Schlieder \tfamily} (of
degree \(\gamma\)). If no such \tfamily\ exists, we will say that \(\Psi\) is
a vector of infinite Reeh-Schlieder degree.
But we note that, by the standard Reeh-Schlieder property
\eqref{eq:RSProp}\footnote{If the Haag-Kastler net under consideration is obtained
  from a suitable Wightman theory (e.g.\ satisfying certain energy bounds
  \cite{Bu90b}), property~\eqref{eq:RSProp} holds as a consequence of the
  original results of Reeh and Schlieder \cite{RS61}.  Alternatively,
  \eqref{eq:RSProp} follows from assuming {\em
  additivity}\/ of the Haag-Kastler net, see e.g.\ \cite{ArQFT99}, Thm.\
  4.14.\label{foot:rsref}}, at least the first inequality of
  \eqref{RS1-section1} can always be satisfied for non-empty regions~\(\Reg \subset
  \RealNum^{\sPOne}\).

  We amend the Haag-Kastler postulates by the following more specific
  assumptions, which can be seen in combination as a sharpened Wigner concept of
  a particle:
\begin{enumerate}[(HK5')]
 \item[(HK5')]
  \label{hk5p}
  In addition to \eqref{eq:sCond}, the relativistic mass operator \(M:=\sqrt{H^2-\Pp^2}\) has an eigenvalue
  \(m\geq 0\). In other words
  \(E_m:=E(H_m)\neq 0\).
\item[(HK6')]
  \label{hk6p}
  The single-particle subspace \(\HilbertSpace_m:=E_m\HilbertSpace\)
  contains a dense subset of vectors of finite Reeh-Schlieder degree.
\end{enumerate}
Under the above assumptions, \eqref{eq:HK1} -- \eqref{eq:HK4}, (HK5'), and (HK6'),
our results from Sections~\ref{sec:conv} and \ref{sec:Fock} below allow to
construct wave-operators and the S-matrix  in
the usual manner (see e.g.\ \cite{Dy09}~App.~A).

\section{Creation operators and their basic properties}
\label{sec:basic}
Given a single-atom state~\(\Psi_1 \in E(H_m)\HilbertSpace\) of mass \(m \geq 0\)
we now want to find a corresponding \tfamily\ of creation operators \(\BB_\T\),
which is suitable for the construction of scattering states.
By the Reeh-Schlieder property \eqref{eq:RSProp} we can always fix some
non-empty bounded open region \(\Reg \subset \RealNum^{\sPOne}\) and pick a
corresponding \tfamily\ of local operators \((\Aa_{\beta})_{\beta>0} \subset
\Alg(\Reg)\) as in formula~(\ref{RS}).

The Klein-Gordon equation will provide a free reference dynamics for
comparison to the large-\({\T}\) asymptotics of the translated operator
\tfamily\ \(A_\beta(\T, \vec x) := U(\T, \vec x) A_\beta U(\T, \vec x)^*\), 
\(x = (\T, \vec x) \in \RealNum^{\sPOne}\), \(\beta > 0\), when taking the
simultaneous limit \(\beta \rightarrow 0\). We will say that \(f :
\RealNum^{\sPOne}
\longrightarrow \ComplexNum\) is a \term{regular positive-energy Klein-Gordon
solution} (of \term{mass} \(m \geq 0\)) if it can be written as 
\[
  f(t, \vec x) = \int \frac{\DInt[\s] k}{(2\pi)^3}\;\Ee^{\Ii \vec k \cdot \vec x - \Ii
    \omega_m(\vec k) t} \tilde f(\vec k),
    \quad \omega_m(\vec k) := \sqrt{\vec k^2 + m^2},
    \label{eq:defKG} \numberthis
\]
where the \term{wave-packet} \(\tilde f\) has to be smooth and compactly
supported.  For the case~\(m = 0\) we will also add the standard requirement
\(\vec 0 \not \in \support \tilde f\), as it leads to improved decay in the
interior of the light cone which will be technically convenient in
\Cref{sec:commut}.

Taking a Reeh-Schlieder \tfamily\ \(A_\beta\) for a given single-particle state
\(\Psi \in E(H_m)\HilbertSpace\) of mass \(m \geq 0\) and a regular
positive-energy Klein-Gordon solution \(f\) of the same mass, we may modify the
standard prescription for creation-operator approximants by admitting the
following additional time-dependence of the smeared operators,
\begin{align}
  \AA_\T &
  := \int \DInt[\s] x\, f(\T, \vec x) \Aa_{\beta(\T)}(\T, \vec x). 
\end{align}
For now it will suffice to demand that the {\em scaling function} \(\beta\)
satisfies \(\beta(\T) \longrightarrow 0\) for \(\T \rightarrow \pm
\infty\).\footnote{For concreteness the reader may take
\(\beta(\T) := \abs{\T}^{-\mu}\), with \(\mu > 0\) fixed. We will later see
that this is a suitable choice in the context of Reeh-Schlieder \tfamilies\ of
finite degree.}
The operator \tfamily\ \(\AA_\T\) then already satisfies some properties which are
characteristic for creation operators, as might be expected from the close
similarity to standard Haag-Ruelle theory\footnote{See e.g.\ \cite{Ha58,Ru62},
  \cite{Dy05}, or \cite{ArQFT99}
Ch.~5.}.
Before proceeding we would like to perform some further standard modifications
needed for the multi-particle case, which will lead to improved
differentiability and impose restrictions on energy-momentum transfers (see
  \Cref{prop:elemProp}
\itref{it:emTrafo}).
\begin{Rem}[uniform differentiability of \(A_\beta\)]
  \label{rem:unifDiff}
By a standard smearing argument, restricting
\(A_\beta\) (for fixed \(\beta\)) to the \(*\)-algebra of smooth
operators~\(\Alg_0(\Reg)\), for which \((t,\vec x) \longmapsto \Aa_\beta(t, \vec
x)\) is arbitrarily often 
differentiable in norm, results in no loss of generality. It is important for our
purposes that this smearing argument directly generalizes to yield {\em
uniformly differentiable}\/ \tfamilies, i.e.
\[
  \norm{\partial_\alpha \Aa_\beta} \leq C_{\alpha} 
  \norm{\Aa_\beta}
  \label{eq:unifDiff}\numberthis
\]
for all multi-indices \(\alpha \in \NaturalNum^{\sPOne}_0\) and some
\(\beta\)-independent constants \(C_\alpha\).
In the following we will therefore assume that all appearing Reeh-Schlieder
\tfamilies\ \(A_\beta\) are smooth and uniformly differentiable. 
\end{Rem}

Further it will be convenient to have at hand a related operator \tfamily\ with
common compact energy-momentum transfers disjoint from a neighbourhood of the
origin. To achieve this we have to give up strict localization and
smear the \tfamily~\(\Aa_\beta\) with the Fourier transform of a function 
\(\hat \chi \in \ContinuousFuncs^\infty_c(\RealNum^{\sOne} \setminus \bar V^-)\).
We will denote the resulting \tfamily\ of almost-local\footnote{See
\Cref{app:proofs}.} operators by
\[
  \Bb_\beta : = \Aa_\beta(\chi).
\]
With these preparations we can introduce our family of creation operator
approximants. 
\begin{Def}[creation operator approximant]
  Let \(\Aa_\beta \in \Alg(\Reg)\) be a uniformly differentiable Reeh-Schlieder
  \tfamily\ for \(\Psi_1 \in E(H_m)\HilbertSpace\), \(m \geq 0\).
  Fixing \(\hat \chi \in \ContinuousFuncs^\infty_c(\RealNum^{\sOne} \setminus
  \bar V^-)\)
  we set \(\Bb_\beta := \Aa_\beta(\chi)\) and for \(\T \in \RealNum\) and a
  regular positive-energy Klein-Gordon solution~\(f\) of the same mass~\(m\) we
  define {\em creation-operator approximants} as
\begin{align}
  \BB_\T &
  := \int \DInt[\s] x\, f(\T, \vec x) \Bb_{\beta(\T)}(\T, \vec x). \quad\label{AT-s2}
\end{align}
\end{Def}
We will often make use of the fact that \(\BB_\T\) are related to the simpler
operator \tfamily~\(\AA_\T\) by convolution algebra.
Let us collect the most important properties of these families of operators.
\begin{Prop} [Basic properties of creation operators] 
  \label{prop:elemProp}
  For an arbitrary operator \tfamily\ \(\Aa_\beta \in \BoundedOps(\HilbertSpace)\) define \(\Bb_\beta\),
  \(\AA_\T\) and \(\BB_\T\) as before. Then
  \begin{enumerate}[(i)]
    \item \label{it:conv}
    \( 
      \BB_\T = \AA_\T(\chi) 
    \).
  \item \label{it:naivebound}
    \(\norm{\BB_\T} \leq C \norm{\AA_{\T}}
      \leq C' (1 + \abs {\T}^N) \norm{\Aa_{\beta(\T)}}
    \) with suitable constants \(C, C', N > 0\).
  \item \label{it:emTrafo}
    For any closed \(\Delta \subset \RealNum^{\sPOne}\), we have the
    \emph{energy-momentum transfer relations}
    \begin{align*}
      \Bb_\beta E(\Delta)\HilbertSpace &\subset E(\Delta + \support \hat
  \chi)\HilbertSpace, \\
  \Bb_\beta^* E(\Delta)\HilbertSpace &\subset E(\Delta - \support \hat
    \chi)\HilbertSpace. \end{align*}
  \item \label{it:annihil} 
    There exists a neighbourhood of zero \(\Uu \subset \RealNum^{\sPOne}\)
    such that
    \(\Bb_\beta^* E(\Uu) = 0\).
  \item \label{it:vacAnnihil}\(\Bb_\beta^* \Omega = 0\).
  \item \label{it:singleP} If \(\Aa_\beta \Omega \rightarrow \Psi_1 \in E(H_m)\HilbertSpace \)
    where \(m \geq 0\) denotes the mass of~\(f\), then
    \[
    \lim_{\T\rightarrow \pm\infty} \AA_\T \Omega = \tilde f(\Pp) \Psi_1, \;\; \text{and similarly} \; 
    \lim_{\T\rightarrow \pm\infty} \BB_\T \Omega =  \tilde f(\Pp) \Psi_1',
    \label{eq:convSingle} \numberthis
    \]
    with \(\Psi_1' := \lim\limits_{\beta \rightarrow 0} \Bb_\beta \Omega = (2\pi)^{\sPOneH} \hat \chi(H, \Pp) \Psi_1\).
  \end{enumerate}
  Properties \itref{it:emTrafo}--\itref{it:vacAnnihil} also hold with
  \(\BB_\T\) in place of \(\Bb_\beta\) without further modifications.
  \proof[Proof]
  \itref{it:conv} is equivalent to
\((\alpha_\T(\Aa_{\beta(\T)}(\chi)))\lss(f_\T\rss)=
((\alpha_\tau(\Aa_{\beta(\T)}))\lss(f_\T\rss))(\chi)\),
where \(f_\T(\vec x):= f(\T, \vec x)\), and this follows from convolution algebra.
Property~\itref{it:naivebound} is a consequence of Hölder's inequality
\(\norm{A(f)} \leq \norm{A} \cdot \norm{f}_1\) and the standard polynomial
bounds for spatial \(\LSpace^1\)-norms of Klein-Gordon solutions \cite[Appendix~1 to
XI.3]{RS3}.
For the proof of relation~\itref{it:emTrafo} we refer to the literature of
Arveson spectral theory --- e.g.~\cite{Arv80}.
To establish \itref{it:annihil}, we note that by assumption \(-\support \hat \chi\) is
compact and disjoint from the closed set \(\bar V^+\), so that for a
sufficiently small neighbourhood \(\Uu\) of the origin there holds 
\((\Uu - \support \hat \chi) \cap \bar V^+ = \emptyset\).
By \itref{it:emTrafo} and the spectrum condition~\eqref{eq:sCond} it follows
that \(\Bb_\beta^* E(\Uu)\HilbertSpace
\in E(\Uu - \support \hat \chi)\HilbertSpace = \{0\}\).
Identity~\itref{it:vacAnnihil} is a direct consequence of
\itref{it:annihil}, as \(\Omega \in E(\Uu)\HilbertSpace\) for any neighbourhood
of zero \(\Uu\). The relations for \(\BB_\T\) follow by similar argument after 
using identity~\itref{it:conv}.

It remains to verify that \(\AA_\T\) and \(\BB_\T\) provide solutions for the
single-particle problem~\itref{it:singleP}.
By spectral calculus we obtain
\[
  \AA_\T\Omega = 
  \tilde f_\T(\Pp) U(\T) \Aa_{\beta(\T)}\Omega 
  = \tilde f(\Pp) \Ee^{ \Ii(H-\omega_m(\Pp)) \T} \Aa_{\beta(\T)}\Omega.
\]
As \(\Psi_1\) is invariant under the unitaries \(V(\T) : = \Ee^{
\Ii(H-\omega_m(\Pp)) \T}\) we may directly estimate 
\begin{align*}
  \normm{\AA_\T\Omega - \tilde f(\Pp) \Psi_1}
  &= \normm{\AA_\T\Omega - \tilde f(\Pp) V(\T) \Psi_1}
   \leq \normm{\tilde f}_\infty \norm{\Aa_{\beta(\T)}\Omega - \Psi_1}.
   \end{align*}
The convergence of \(\BB_\T\Omega\) follows then from \itref{it:conv} by writing
\(
  \BB_\T\Omega = (2\pi)^{\sPOneH} \hat \chi(H, \Pp) \AA_\T\Omega 
\). \qed
\end{Prop}

An important consequence of the energy-momentum transfer
relation~\itref{it:emTrafo} is the following energy bound.
The key point is that the estimate can be made uniform in \(\T\)
relative to the norm of the underlying Reeh-Schlieder \tfamilies, as long as we
consider the restriction of creation operators to a subspace of bounded energy.
Our analysis was somewhat inspired by Herdegen's work \cite{Hdg13}, but we rely
on different aspects of Buchholz' results~\cite{Bu90} given in \Cref{lem:bu22}.

\begin{Lem}[\cite{Bu90}, Lemma 2.2] \label{lem:bu22}
  Let \(K \subset \RealNum^{\s}\) compact, \(\Bb \in
  \BoundedOps(\HilbertSpace)\) and denote by \(P_n\) the orthogonal projection
  onto the intersection of the kernels of the \(n\)-fold products of translated
  operators \(\Bb(\vec x_1)\ldots \Bb(\vec x_n)\) for any configuration of \(\vec
  x_1, \ldots, \vec x_n \in \RealNum^{\s}\). Then
  \[
    \norm{P_n \int\limits_K \DInt[\s] x \; (\Bb^*\Bb)(\vec x) P_n}
    \leq (n-1) \int\limits_{\Delta K} \DInt[\s] x
    \norm{[\Bb^*, \Bb(\vec x)]},
  \]
  where integration on the right is over all element-wise differences 
  \(\Delta K := K-K\).
\end{Lem}

\begin{Prop}[Energy bounds] \label{lem:bu} 
  Without further restrictions on the \tfamilies\ of operators \(\Aa_\beta\),
  \(\Aa_{k\Oc\beta} \in \Alg(\Reg)\), we have for any compact \(\Delta \subset
  \RealNum^{\sPOne}\),
  \begin{align*}
    \norm{\BB_\T E(\Delta)} &\leq C \norm{\Aa_{\beta(\T)} },
    \label{eq:bu1} \numberthis
    \\
    \norm{\BB_{1\Oc\T_1} \ldots \BB_{n\Oc\T_n} E(\Delta)} &\leq C
    \prod_{k=1}^n \norm{\Aa_{k\Oc\beta(\T_k)}},
    \label{eq:buN} \numberthis
  \end{align*}
  where the constant \(C\) depends on \(\Delta\), \(\Reg\), \(\support \hat
  \chi\), the number of operators \(n\), and the corresponding wave packets
  \(\tilde f\),
  \(\tilde f_k\), but it is independent of \(\T\).
 \proof
  To establish \eqref{eq:bu1}, let \(\Delta \subset \RealNum^{\sPOne}\) be a
  given compact set.  By a partition argument, we can assume that \(\support
  \hat \chi\) is contained in a compact, convex set disjoint from \(\bar V^-\).
  The compact common ener\-gy-\-mo\-men\-tum transfer 
  (cf.~\Cref{prop:elemProp} \itref{it:emTrafo}) of \(\BB_\T\) then
  allows us to write
  \[
    \norm{\BB_\T E(\Delta)} = \norm{E(\Delta + \support \hat \chi) \BB_\T E(\Delta)}
      \leq \norm{E(\Delta') \BB_\T} = \norm{\BB_\T^* E(\Delta')}, 
  \]
  where \(\Delta' : = \Delta + \support \hat \chi\) is compact as well.

  To make the connection with \Cref{lem:bu22}, we note that by iterated
  application of \Cref{prop:elemProp}~\itref{it:emTrafo} and
  translation-invariance of finite-energy subspaces, we obtain 
  \[ 
    \Bb_\beta^*(\vec x_1) \ldots \Bb_\beta^*(\vec x_n) E(\Delta) \HilbertSpace
      \subset E(\Delta - \Sigma_n \support \hat \chi)\HilbertSpace,
  \]
  where 
  \(\Sigma_n \support \hat \chi := \{ y_1 + \ldots + y_n: y_k \in \support \hat \chi\}
  = n \support \hat \chi \) due to convexity.  By the Hyperplane Separation
  Theorem, we obtain \((\Delta' - \Sigma_n \support \hat \chi )\cap \bar V^+ =
  \emptyset\) for  sufficiently large \(n \in \NaturalNum\).
  This implies via the spectrum condition~\eqref{eq:sCond} that for such \(n\),
  the projections \(P_n\) appearing in \Cref{lem:bu22} may be estimated from
  below by
  \(
    E(\Delta')\HilbertSpace \subset P_n \HilbertSpace. 
  \) 
  With these preparations we can estimate 
  \begin{align*}
    \norm{\BB_\T^*E(\Delta')} &\leq \norm{\BB_\T^*P_n}
    \leq \sup_{\substack{\Psi \in \HilbertSpace\\\norm{\Psi} = 1}}
    \int \DInt[\s] x 
    \abs{f(\T, \vec x)} \norm{\Bb_{\beta(\T)}^*(\T, \vec x) P_n \Psi}
    \\&\leq
    \left(\int \DInt[\s] x \abs{f(\T, \vec x)}^2\right)^{1/2}
    \left(\sup_{\substack{\Psi \in \HilbertSpace\\\norm{\Psi} = 1}}
      \int \DInt[\s] x  \norm{\Bb_{\beta(\T)}^*(\T, \vec x) P_n \Psi}^2
    \right)^{1/2}.
  \end{align*}
  The first factor is constant by the Plancherel identity (cf.\
  Prop.~\ref{lem:decayWave} \itr{iv}).
  For estimating the second factor we
  choose an arbitrarily large compact region \(K \subset \RealNum^\s\) and
  obtain from \Cref{lem:bu22} that
  \begin{align*}
\sup_{\substack{\Psi \in \HilbertSpace\\\norm{\Psi} = 1}} \;
      \int\limits_K \DInt[\s] x  \norm{\Bb_{\beta(\T)}^*(\T, \vec x) P_n \Psi}^2
  & = 
\sup_{\substack{\Psi \in \HilbertSpace\\\norm{\Psi} = 1}}
  \left \langle \Psi , P_n \int\limits_K \DInt[\s] x (\Bb_{\beta(\T)}\Bb_{\beta(\T)}^*)(\T, \vec x) P_n
  \Psi \right\rangle 
  \\&=\norm{P_n \int\limits_K \DInt[\s] x (\Bb_{\beta(\T)}\Bb_{\beta(\T)}^*)(\T, \vec x) P_n}
  \\&\leq (n-1) \int\limits_{\Delta K} \DInt[\s] x \norm{\left[\Bb_{\beta(\T)}, \Bb_{\beta(\T)}^*(\vec
x)\right]}. 
  \end{align*}
  The \tfamily\ \(B_\beta\) and its adjoint
  are uniformly almost-local (as defined in \Cref{app:proofs}), so that the remaining integral can 
  be estimated by \( 2 C_\chi \normm{\Aa_{\beta(\T)}}^2\cdot d^\s\), where \(d\)
  depends only on the size of the localization region of \(\Aa_\beta\). 
  This yields a bound which is uniform in \(\Delta K\) and by taking \(K \nearrow
  \RealNum^\s\) we obtain the energy bound for a single operator.
 
  Then the bound \eqref{eq:buN} on multiple creation operators follows directly by
  induction: the compact common energy-momentum transfer of the
  family \(\BB_{k\Oc\T}\) yields
  \begin{align*}
    \norm{\BB_{1\Oc\T_1} \ldots \BB_{n\Oc\T_n} E(\Delta)} &=
    \norm{\BB_{1\Oc\T_1} \ldots \BB_{n-1\NOc\T_{n-1}} E(\Delta + \support \hat \chi)
    \BB_{n\Oc\T_n} E(\Delta)} 
    \\&\leq
    \norm{\BB_{1\Oc\T_1} \ldots \BB_{n-1\NOc\T_{n-1}} E(\Delta + \support \hat \chi)}
    \cdot\norm{ \BB_{n\Oc\T_n} E(\Delta)} 
    \\&\leq C^{(n-1)}_{\Delta + \support \hat \chi} \left(\;\prod_{k=1}^{n-1}
    \norm{\Aa_{k\Oc\beta(\T_k)}}\,\right) \cdot  C_\Delta \norm{\Aa_{n\beta(\T_n)}}. \qedhere
  \end{align*}
\end{Prop}

\section{Geometry of non-equal time commutators}
\label{sec:commut} 

The goal of this section is to study the decay behaviour of commutators 
\( \left[ \BB_{1\Oc\T_1}, \BB_{2\Oc\T_2} \right] \) for distinct asymptotic
parameters \(\tau_1 \not = \tau_2\). The strongest known decay estimates for
equal times \(\tau_1 = \tau_2\) have been established for the case, where
the defining Klein-Gordon solutions~\(f_1\), \(f_2\) have disjoint support in
momentum space~\cite{Hep65}.
This corresponds to the physically reasonable assumption that the two particles
will separate at large times. We will restrict our analysis to this
setting and begin by reviewing required results on regular Klein-Gordon
solutions  \(f: \RealNum^{\sPOne} \longrightarrow \ComplexNum\) with mass \(m
\geq 0\), as defined in~\eqref{eq:defKG}.

The geometry of the asymptotic behaviour of \(f\) can be intuitively understood
in terms of the set of velocities corresponding to the momenta \(\vec k \in
\support \tilde f\). Accordingly we define the \term{velocity support} of \(f\)
by \( \vec \Gamma_{\tilde f} := \{  \vec k/\omega_m(\vec k) \in \RealNum^\s: \; \vec k \in
    \support \tilde f\} \). 
Let us recall how this definition allows for a compact formulation of the
classical result of Ruelle~\cite{Ru62} on the decay of Klein-Gordon solutions
outside the velocity-support cone. We provide a unified treatment of the massive
and massless case.

\begin{Lem}[velocity-support estimate] \label{lem:velo}
 Let \(f\) be a regular solution of the Klein-Gordon equation with mass \(m
  \geq 0\).  The following estimate holds for any \(N \in \NaturalNum\) 
 with suitable constants \(C_N > 0\) and any \((t, \vec x)\in \RealNum^{\sPOne}\)
 satisfying \( \vec x/t \not \in
 \vec\Gamma_{\tilde f}\),
  \begin{align*}
    \abs{f(t, \vec x)} \leq \frac{C_N}{\delta^N \abs{t}^N},
  \end{align*}
  where \(\delta\) denotes the distance of \(\vec x/t\) from the set \(\vec
  \Gamma_{\tilde f}\).
\end{Lem}
\noindent
For regular massive Klein-Gordon solutions, geometrical propagation properties
such as the above can be found in various textbooks, e.g.\ \cite{ArQFT99} Thm.\
5.3.
We will skip the standard proof, which makes use of the non-stationary phase
method (see e.g.\ \cite{RS3}, Appendix 1 to XI.3). 
\Cref{lem:velo} is applicable in particular in the 
case of \(\vec x/t\) approaching the velocity support \(\vec \Gamma_{\tilde
f}\). This will be needed later in \Cref{lem:decayWave} to establish certain
norm estimates in the
massless case.

For the purpose of rapid decay of commutators, it is actually
sufficient to make use of \Cref{lem:velo} in some fixed neighbourhood \(\vec U
\supset \vec \Gamma_{\tilde f}\). One obtains the following simple rapid-decay
estimate with respect to time {\em and}\, space outside a corresponding enlarged
neighbourhood of the cone generated by the velocity support.

\begin{Cor}
  \label{cor:asympt}
  Let \(f\) be a regular solution of the Klein-Gordon equation with mass \(m
  \geq 0\) and let \(\vec U \supset \vec \Gamma_{\tilde f}\) be any (slightly
  larger) neighbourhood of the velocity support.
  Then the restriction of \(f\) to the complement of the cone
  \[
    \Upsilon_{\vec U} := \{ (t, t \vec v) \in \RealNum^{\sPOne}, \vec v \in \vec U, t \in
    \RealNum\}
  \]
  is rapidly decreasing, i.e.\ for any \(N \in \NaturalNum\) we have
  \[
    \abs{f(t, \vec x)} \leq {C_{N}}(1+\abs{t}+ \abs{\vec x})^{-N}
  \qquad
  \forall \, (t, \vec x) \in \RealNum^{\sPOne}\setminus \Upsilon_{\vec U},
  \]
  with suitable \(C_N > 0\) depending on \(N\), \(\tilde f\), and the distance between \(\RealNum^\s
  \setminus \vec U\) and \(\vec \Gamma_{\tilde f}\).  \end{Cor}

\begin{figure}
  \begin{center}
\def\sca{0.7}
\begin{tikzpicture}[scale=\sca]
  \usetikzlibrary{calc}
\def\gal{-0.7} 
\def\gar{-0.4} 

\def\gbl{0.3} 
\def\gbr{0.6} 

\def\r{1.4} 
\def\ta{6} 
\def\tb{7.1} 

\def\ex{0.1} 
\def\dt{0.4} 
\def\cc{6} 

\def\un{3} 
\def\eex{0.45} 

\draw[fill=gray!60] (0,0) circle (\eex);
\draw (0.1,-0.4) node[below right=0.0] {$\Reg_{\Aa_{k\Oc\beta}}$};

\draw[fill=blue!5,below] (0,0) -- (9*\gar,9) -- (9*\gal,9) -- (0,0) ;
\draw[fill=blue!5,below] (0,0) -- (9*\gbr,9) -- (9*\gbl,9) -- (0,0) ;
\draw (9*\gal/2+9*\gar/2,9.0) node[above]{ $\Upsilon_{\vec U_1}$};
\draw (9*\gbl/2+9*\gbr/2,9.0) node[above]{ $\Upsilon_{\vec U_2}$};

\draw[rounded corners=14*\eex/\sca, fill=gray!80, fill opacity=0.5] ({\ta*(\gal+\gar)/2},\ta+\eex) -- (\ta*\gar+\eex,\ta+\eex)
-- (\ta*\gar+\eex,\ta-\eex)
-- (\ta*\gal+\eex,\ta-\eex)
-- (\ta*\gal-\eex,\ta-\eex)
-- (\ta*\gal-\eex,\ta+\eex)
-- ({\ta*(\gal+\gar)/2},\ta+\eex);

\draw (\ta*\gal,\ta) node[left=1,anchor=north east]{${\AA_{1\Oc\T_1}^\uparrow}$};

\draw[rounded corners=14*\eex/\sca, fill=gray!80, fill opacity=0.5] ({\tb*(\gbl+\gbr)/2},\tb+\eex) -- (\tb*\gbr+\eex,\tb+\eex)
-- (\tb*\gbr+\eex,\tb-\eex)
-- (\tb*\gbl+\eex,\tb-\eex)
-- (\tb*\gbl-\eex,\tb-\eex)
-- (\tb*\gbl-\eex,\tb+\eex)
-- ({\tb*(\gbl+\gbr)/2},\tb+\eex);

\draw (\tb*\gbr,\tb) node[below=7,left=6,anchor=north west]{${\AA_{2\Oc\T_2}^\uparrow}$};

\draw[|-|] 
   (0,\un+0.01) 
  -- (0,\un-0.01) node[left=-0.4,anchor=west] {\footnotesize $1$};

\def\rs{5.9}
\draw[|-|] 
  (\rs,\ta+\r)node[right=-0.4,anchor=west] {\footnotesize $(1\! +\! \rho) \T_1$} 
  -- (\rs,\ta) node[left=-0.4,anchor=west] {\footnotesize $\T_1$};
\draw[|-|] 
   (0,\tb+0.01) 
  -- (0,\tb-0.01) node[left=-0.4,anchor=west] {\footnotesize $\T_2$};
\draw[|-|] 
   (\rs,\ta+0.01) 
   -- (\rs,\ta-\r)node[right=-0.4,anchor=west] {\footnotesize $(1\! -\! \rho) \T_1$};

\draw[thick,|-|] (\ta*\gal,\ta) -- (\ta*\gar,\ta);
\draw[thick,|-|] (\tb*\gbl,\tb) -- (\tb*\gbr,\tb);

\def\usp{0.3}
\def\uusp{0.02}

\def\eps{0.020}
\draw[thick,|-|] ({(\gal+\eps)*\un},\un)--({(\gar-\eps)*\un},\un);
\draw[thick,|-|] ({(\gbl+\eps)*\un},\un)--({(\gbr-\eps)*\un},\un);

\def\uun{3.6}
\draw[thick,|-|] ({(\gar-1.5*\eps)*\un},\uun) -- node[above] {$d$\;\;}
  ({(\gbl+1.5*\eps)*\un},\uun);

\draw ({(\gal + \gar)*\un/2}, \un) node[below,anchor=north east] {\;$\vec
\Gamma_{\!\tilde f_1}$};
\draw ({(\gbl + \gbr)*\un/2}, \un) node[below,anchor=north west] {\;\,$\vec
  \Gamma_{\!\tilde f_2}$};

\draw (0, \ta) coordinate (a);
\draw[thick] (a){}+(3.2,3.2) -- ++(-3.5,-3.5); 
\draw[thick] (a){}+(-3.2,3.2) -- ++(3.5,-3.5); 

\draw[->] (-7,0) -- (7,0);
\draw (7.5,0) node {$\vec x$};
\draw[->] (0,-1) -- (0,9.5);
\draw (0,10) node {$t$};
\end{tikzpicture}
\end{center}
\vspace*{-1.5em}
\caption{Localization regions of asymptotically dominant parts \(\AA_{k\Oc
  \T_k}^\uparrow\) with disjoint velocity supports and \(\tau_1 \not = \tau_2\)
  (schematically; a separating pair of wedges is indicated, restricting
\(\abs{\T_2 - \T_1}\)).}
\label{fig:loc}
\end{figure}
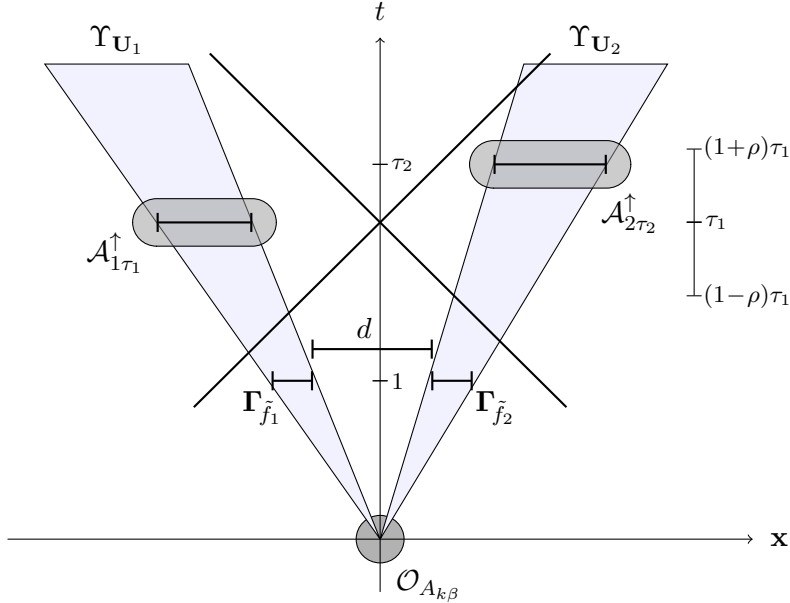

While our construction of collision states will make use of the creation
operators~\(\BB_{k\Oc\T}\), it is clear that additional technical difficulties
arise due to the loss of strict locality when passing from localized Reeh-Schlieder
\tfamilies~\(\Aa_{k\Oc\beta} \in \Alg(\Reg)\) (with \(\Reg\) independent of~\(\beta\))
to the almost-local operators~\(\Bb_{k\Oc\beta}:= \Aa_{k\Oc\beta}(\chi)\).
We recall that the thus obtained compact energy-momentum transfers of
\(\Bb_{k\Oc\beta}\) were essential for establishing energy bounds in
\Cref{lem:bu}.

One strategy to resolve these complications, which makes arguments based on locality
particularly transparent, is to first establish corresponding results for
the operators~\(\AA_{k\Oc\T}\), as these have better localization
properties. Statements which are sufficiently stable under smearing can
then be carried over to \(\BB_{k\Oc\T} = \AA_{k\Oc\T}(\chi)\) (see
\Cref{prop:elemProp}~\itref{it:conv}). For this reason
we want to additionally allow space-time
translates \(\alpha_x(\BB_\tau)\) with \(x \in \RealNum^{\sPOne}\) restricted to
suitable bounded regions in space-time.  We note for clarification that
\(\alpha_x(\AA_{k\NOc\T+t})\) differs from \(\alpha_{x+(t,0)}(\AA_{k\Oc\T})\)
due to the time evolution of the Klein-Gordon solution and the underlying
time-dependent Reeh-Schlieder \tfamily.

The geometrical content of \Cref{lem:estNoneqB} is illustrated in
\Cref{fig:loc}. Regarding the depicted situation it is clear that in order to
obtain rapid decay the allowed translation vectors~\(x = (x^0 , \vec x) \in
\RealNum^{\sOne}\) will have to be subjected to
a similar restriction as the time differences~\(\abs{\T_2-\T_1}\).
In the context of causal distance estimates, it will be convenient to specify
this restriction by introducing the norm~\( \abs{x}_c := \abss{x^0} + \abs{\vec
  x}\), where \(\abs{\vec x} := \sqrt{\vec x^2}\) denotes the Euclidean length
of \(\vec x \in \RealNum^\s\).
The centered open balls generated by this norm are the familiar double 
cones~\(\DoubleCone_R = \{ x \in \RealNum^{\sPOne}: \abs{x}_c < R \}\)
with radius \(R > 0\).

\def\Tmin{\T_{\text{\rm min}}}
\def\brho{\bar \rho}
\begin{Lem}
  \label{lem:estNoneqB}
  There exists a constant \(C > 0\), such that for any \(f_1, f_2\) with
  velocity supports separated by a positive distance \(d > 0\), the following
  estimate holds for any \(N \in \NaturalNum\), \(x \in \RealNum^{\sPOne}\) and
  \(\T_1, \T_2 \in \RealNum\) satisfying \( \abs{x}_c + \abs{\T_2-\T_1} \leq C
  d^2 \cdot \Tmin\),
  \[
    \norm{[\AA_{1\Oc\T_1}, \alpha_x(\AA_{2\Oc\T_2})]} \leq 
    {C_N
    \norm{\Aa_{1\Oc\beta(\T_1)}}\norm{\Aa_{2\Oc\beta(\T_2)}}}\cdot{(1 + \Tmin)^{-N}}.
    \label{eq:Bcomm} \numberthis
  \]
  Here, \(\Tmin := \min(\abs{\T_1}, \abs{\T_2})\) and the constants \(C_N\)
  depend only on \(N\), \(f_k\) and the size of the localization regions of
  \(\Aa_{k\beta}\). 
  \proof  
  We can assume without restriction that \(\Tmin = \abs{\T_1}\).
  Further it is enough to establish \eqref{eq:Bcomm} for \(\abs{\T_1}\)
  sufficiently large\footnote{On any bounded interval \(\abs{\T_k} \leq
    {\T_{\text{max}}}\) (\(\T_{\text{max}}\) fixed), we may use
    \Cref{prop:elemProp}~\itref{it:naivebound} to obtain \(
      \norm{[\AA_{1\Oc\T_1}, \alpha_x(\AA_{2\Oc\T_2})]} \leq C_{\T_{\text{max}}}
  {\norm{\Aa_{1\Oc\beta(\T_1)}}\norm{\Aa_{2\Oc\beta(\T_2)}}}\),
  which is compatible with \eqref{eq:Bcomm} for sufficiently large~\(C_N\).},
  and for this case we will make use of a suitable common asymptotic decomposition of
  the Klein-Gordon solutions~\(f_k\). 
  By definition, the corresponding velocity supports \(\vec \Gamma_{\tilde f_1}\) and
  \(\vec \Gamma_{\tilde f_2}\) are closed subsets of the closed unit ball.
  Aiming at the application of
  \Cref{cor:asympt}, it is clear that we can find neighbourhoods \(\vec U_{1}\)
  and \(\vec U_2\) of the velocity supports \(\vec \Gamma_{\tilde f_1}\) and
  \(\vec \Gamma_{\tilde f_2}\), which are separated by a distance of at least
  \(d/2\) and which are contained in some fixed larger ball. For concreteness we
  may assume without loss of generality that \(\vec v \in \vec
    U_{1/2}\) always satisfy\footnote{Such a bound will be important later in
    the proof.  
    The concrete choice of the constant has no physical significance, but it
    will influence the magnitude of the proportionality constant \(C\) 
    controlling time-differences in the statement of the lemma.
  }
    \(\abs{\vec v} \leq 2\).

  Denoting by \(\CharFct_{\Upsilon_{\vec U_k}}\) the characteristic function
  of the cone \(\Upsilon_{\vec U_k}\) (as defined in \Cref{cor:asympt}) 
  we introduce the following decompositions into \term{asymptotically dominant}
  and \term{negligible parts},
  \[
    f_k = f_k^\uparrow + f_k^\downarrow,
    \quad f_k^\uparrow(x) := f_k(x) \cdot \CharFct_{\Upsilon_{\vec U_k}}(x),
  \]
  and similarly \(\AA_{k\Oc\T} = \AA_{k\Oc\T}^\uparrow +
  \AA_{k\Oc\T}^\downarrow\), (\(k=1,2\)), denote the induced decompositions of creation operators.
  By \Cref{cor:asympt}, we obtain
  \[
    \norm{\AA_{k\Oc\T_k} - \AA_{k\Oc\T_k}^\uparrow}
    = \norm{\AA_{k\Oc\T_k}^\downarrow}
    \leq {C_N' \norm{\Aa_{k\Oc\beta(\T_k)}}}\cdot ({1+\abs{\T_k}})^{-N}.
  \]
  This implies that it is sufficient to analyse the commutator of the
  dominant parts as can be seen from the following estimate,
  which holds uniformly in \(x\in\RealNum^{\sPOne}\),
  \[
    \norm{[\AA_{1\Oc\T_1}^{\phantom{\uparrow}}, \AA_{2\Oc\T_2}(x)]} \leq 
    \norm{[\AA_{1\Oc\T_1}^\uparrow, \AA_{2\Oc\T_2}^\uparrow(x)]} 
    + {C_N'' \norm{\Aa_{1\Oc\beta(\T_1)}}\norm{\Aa_{2\Oc\beta(\T_2)}}}
    \cdot (1 + \Tmin)^{-N}.
  \]

  We will now verify that the commutator of the dominant parts vanishes for
  sufficiently large \(\T_1\) in the claimed region of~\(x\) and \(\T_k\). 
  As a standard consequence of the Haag-Kastler axioms we obtain
  \[
    \AA_{k\Oc\T_k}^\uparrow \in \Alg(\Reg_{k,\T_k}),
  \qquad\text{with} \;\Reg_{k,\T_k} := \DoubleCone_R + \T_k\cdot (\{1\} \times
  \vec U_k),
  \]
  where we picked a sufficiently large radius \(R > 0 \) such that the double
  cone~\(\DoubleCone_R\) provides a common bounded
  localization region of the \tfamilies~\(\Aa_{k\Oc\beta}\). 
  Then we have by covariance \( \AA_{2\Oc\T_2}^\uparrow(x) \in \Alg(\Reg_{2,\T_2}+x)\).
  To estimate the causal distance of any two points \(y_1 \in \Reg_{1,\T_1}\) and
  \(y_2 \in \Reg_{2,\T_2}+x\) from the respective support regions, we write them
  as \(y_1 = o_1 + \T_1 \cdot (1,\vec v_1)\), \(y_2 = o_2 + \T_2 \cdot(1, \vec
  v_2) + x\), with \(o_1, o_2 \in \DoubleCone_R\) and \(\vec v_k \in \vec U_k\).
  We can then see that
  \begin{align*}
    y_2 - y_1 = [(\T_2, \T_2 \vec v_2) - (\T_1, \T_1 \vec v_1)] +  o_2 + x - o_1.
  \end{align*}
  In the end we will impose a suitable restriction on \(u := o_2 + x - o_1\) and
  therefore the space-like separation of \(y_1\) and \(y_2\) needs to be derived
  from the difference term inside the brackets, which we denote by \(w := (\T_2,
  \T_2 \vec v_2) - (\T_1, \T_1 \vec v_1)\). We compute 
  \begin{align*}
    w^2 &= (\T_2-\T_1)^2 - (\T_2 \vec v_2 - \T_1 \vec v_1)^2,
    \\
    \abs{\T_2 \vec v_2 - \T_1 \vec v_1}
    & = \abs{\T_2 \vec v_2 - \T_1 \vec v_2 + \T_1 (\vec v_2  - \vec v_1)}
    \\&\geq  - \abs{\T_1-\T_2}  \abs{\vec v_2} 
    + \abs{\T_1} \abs{\vec v_2- \vec v_1},
    \label{eq:tvEst} \numberthis
  \end{align*}
    and thus
  \[
    w^2 \leq -  \T_1^2(\vec v_2- \vec v_1)^2 
    + 2\abs{\T_1-\T_2} \abs{\T_1} \abs{\vec v_2 - \vec v_1} \abs{\vec v_2}
    +(\T_1-\T_2)^2.
  \]
  We note that by the non-vanishing negative coefficient of the quadratic term,
  \(w\) will become space-like for large enough \(\abs{\T_1}\) if sufficient
  restrictions are placed on \(\abs{\T_2 - \T_1}\).  
  By a similar argument also the perturbation of adding \(u\) can be controlled,
  as can be seen from 
  \[
    (y_2 - y_1)^2 = w^2 + 2 w \cdot u + u^2
    \leq w^2 + 2 \abs{w}_c \abs{u}_c + \abs{u}_c^2,
    \label{eq:yest}\numberthis
  \]
  where we used the Cauchy-Schwarz inequality.
  Now assume that \(\abs{\T_2 - \T_1} + \abs{x}_c \leq \brho \abs{\T_1}\) for some
  constant \(\brho > 0\) (to be determined).
  Using that our choice of \(\vec U_k\) implies
  \(\abs{\vec v_k} \leq 2\),
  \(0 < d \leq \abs{\vec v_2 - \vec v_1} \leq 4\),
  we can then further estimate 
  \begin{align*}
    w^2 & \leq - d^2 \T_1^2 + (16\brho+\brho^2)\T_1^2,
    \\
    \abs{w}_c & \leq 3 \abs{\T_2 - \T_1} + 4 \abs{\T_1} \leq  (4+3\brho)
    \abs{\T_1},
    \\
    \abs{u}_c&\leq \abs{o_1}_c + \abs{o_2}_c + \abs{x}_c \leq 2R+\brho
    \abs{\T_1}.
  \end{align*}

  To simplify the resulting bound on \((y_2-y_1)^2\), let us choose firstly 
  \(\brho \leq 1\) and then subsequently \(\abs{\T_1} \geq 2 R/\brho\). This
  allows us to eliminate unimportant scales by writing
  \(\abs{w}_c \leq 7 \abs{\T_1}\),  \(\abs{u}_c \leq 2\brho \abs{\T_1}\) and
  \(\brho^2 \leq \brho\). Then we obtain from \eqref{eq:yest} that
 with a suitable numerical constant~\(C>0\),
  \[
    (y_2-y_1)^2 \leq -d^2 \T_1^2 + C^{-1} \brho \T_1^2.
  \]
This proves that any choice \(0 < \brho < Cd^2 \; (< 1)\) leads to space-like
localization regions of the dominant parts, and so  by locality
\( [ \AA_{1\Oc\T_1}^\uparrow, \AA_{2\Oc\T_2}^\uparrow(x) ] = 0 \) for \(\abs{\T_1} >
2R/\brho\) under the assumed restriction on \(\T_2\) and \(x\). \qed
\end{Lem}

With this technical preparation we can now establish asymptotic commutation of
the creation operators \(\BB_{k\Oc\T}\) with disjoint velocity supports at
non-equal times. We can also appreciate now how the power-law scaling 
\(\beta(\T) = \abs{\T}^{-\mu}\) (for large enough \(\abs{\T}\)), \(\mu > 0\),
plays a distinguished role: for this choice the norm terms
\(\norm{\Aa_{k\beta(\T)}}\) can be absorbed due to the rapid decay in
\Cref{lem:estNoneqB}. While these commutator estimates may still be improved in
a suitably adapted setting, already the results of the next section will impose
sharp restrictions on the scaling parameter \(\mu\).

\begin{Thm}
  [non-equal-time commutator estimate]
  \label{lem:estNoneq}
  Let \(\Aa_{k\Oc\beta}\), (\(k=1,2\)), be Reeh-Schlieder \tfamilies\ of finite degree\footnote{%
    \label{foot:commAssump}%
    For \Cref{lem:estNoneq}, it is sufficient if the operator
    \tfamilies~\(\Aa_{k\Oc\beta}\) are uniformly localized (\(\Aa_{k\Oc\beta}
    \in \Alg(\Reg)\), with bounded \(\Reg\) independent of \(\beta\)) and have
    at most polynomial norm growth \(\norm{\Aa_{k\Oc\beta}} \leq
    \beta^{-\gamma}\), (\(\gamma \geq 0\)).
  }, take regular Klein-Gordon solutions \(f_k\) with disjoint velocity
  supports and assume a fixed scaling \(\beta(\T) = \abs{\T}^{-\mu}\),
  \(\mu > 0\). 
  Setting \(\rho:=Cd^2/2 \in (0,1)\) with \(C,d\) as in \Cref{lem:estNoneqB}, there exists
  for any \(N\in\NaturalNum\) a constant~\(C_N > 0\), such that for
  arbitrary \(\T \in \RealNum\) and all \( \T_{1}, \T_2\) from the corresponding interval
spanned by \(\T\) and \(\T+\rho\T\),
  \[
    \norm{[\BB_{1\Oc\T_1}, \BB_{2\Oc\T_2}]} \leq {C_N} (1 + \abs{\T})^{-N}.
  \]
  \proof We have \(\BB_{k\Oc\T_k} = \AA_{k\Oc\T_k}(\chi)\), with \(\chi \in
  \SchwartzSpace(\RealNum^{\sPOne})\) and so we obtain 
  \begin{align*}
    \norm{\left[\BB_{1\Oc\T_1}, \BB_{2\Oc\T_2}\right]}
      &\leq
      \int \DInt[\sOne]x \,\DInt[\sOne] y \;
      \abs{\chi(x)} \abs{\chi(y)} \norm{\left[\AA_{1\Oc\T_1},\AA_{2\Oc\T_2}(y-x) \right]}.
      \label{eq:bcommint}\numberthis
  \end{align*}
  We decompose the integral into the region \(\abs{x}_c \leq
  \rho \abs{\T}/2\) and its complement, and similarly for the \(y\)-integration.
  As a consequence of our assumptions we have a polynomial bound
  \(\norm{\Aa_{k\Oc\beta(\T)}} \leq \abs{\T}^{\mu\gamma}\) and restricting to
  \(\T_1,\T_2\) from the claimed interval we obtain for fixed 
  \(x \in \RealNum^{\sPOne}\) that
  \begin{align*}
    \int \DInt[\sOne] y \, \abs{\chi(y)} \norm{\left[\AA_{1\Oc\T_1},\AA_{2\Oc\T_2}(x-y) \right]}
    &\leq 2 \norm{\chi}_1 \norm{f_{1\Oc\T_1}}_1 \norm{f_{2\Oc\T_2}}_1
    \norm{\Aa_{1\Oc\beta(\T_1)}}\norm{\Aa_{2\Oc\beta(\T_2)}}
    \leq C \abs{\T}^M,
  \end{align*}
  for some large enough \(M > 0\) and the estimate holds uniformly in \(x\).
  This now implies that the integral of \eqref{eq:bcommint} restricted to the
  outside region \(\abs{x}_c \geq \rho\abs{\T}/2\) is rapidly decreasing: we can
  estimate it by a product of the above polynomially bounded function with
  the rapidly decreasing function obtained by integrating \(\abs{\chi(x)}\)
  over the retracting regions given by \(\abs{x}_c \geq \rho\abs{\T}/2\).
By a similar argument we can assume that also \(\abs{y}_c \leq
\rho\abs{\T}/2\) and so we can write with suitable constants~\(C_N'\),
  \begin{align*}
    \norm{\left[\BB_{1\Oc\T_1}, \BB_{2\Oc\T_2}\right]}
    &\leq \frac{C_N'}{1+\abs{\T}^N} \quad+\quad 
    \int\limits_{\mathclap{ \abs{x}_c , \abs{y}_c\leq {\rho\abs{\T}}/2}}
      \DInt[\sOne]x \,\DInt[\sOne] y \;
      \abs{\chi(x)} \abs{\chi(y)} \norm{\left[\AA_{1\Oc\T_1},\AA_{2\Oc\T_2}(x-y) \right]}.
  \end{align*}
  Assuming the given restriction \(\abs{\T_1 - \T_2} \leq \rho\abs{\T} \;(\leq
  \rho\Tmin)\) we obtain \(\abs{\T_2-\T_1} + \abs{x-y}_c \leq 2 \rho \abs{\T}
  \leq Cd^2 \Tmin\). Therefore \Cref{lem:estNoneqB} is applicable, which yields
  \begin{align*}
    \int\limits_{\mathclap{ \abs{x}_c , \abs{y}_c\leq {\rho\abs{\T}}/2}}
      \DInt[\sOne]x \,\DInt[\sOne] y \;
      \abs{\chi(x)} \abs{\chi(y)} \norm{\left[\AA_{1\Oc\T_1},\AA_{2\Oc\T_2}(x-y) \right]}
      \leq \frac{C_{N'}''
      \norm{\Aa_{1\Oc\beta(\T_1)}}\norm{\Aa_{2\Oc\beta(\T_2)}}}{(1+\Tmin)^{N'}}.
  \end{align*}
  As \(\Tmin \geq \abs{\T}\) we can proceed similarly as before and choose \(N'\)
  large enough (depending on the desired decay order~\(N\), the scaling \(\mu\),
  and \(\norm{\Aa_{k\Oc\beta(\T_k)}}\)) to compensate for the polynomial growth
  of \(\norm{\Aa_{k\Oc\beta(\T_k)}}\).
   \qed
 \end{Thm}

It is clear that the same reasoning applies, if we replace one or more creation operator
approximants by their adjoints.
For later use in \Cref{sec:Fock}, we also mention the following equal-time
result regarding double commutators with one additional creation operator which
may have arbitrary velocity support. This follows from \Cref{lem:estNoneq} by a
well-known decomposition argument.
\begin{Cor}[decay of double commutators]
  \label{lem:doubleComm}
  In the setting of \Cref{lem:estNoneq} let \(\BB_\T\)
  be an additional creation operator approximant defined in terms of a regular
  Klein-Gordon solution~\(f\) (without restrictions on its velocity support) and
  an additional Reeh-Schlieder \tfamily~\(\Aa_\beta\) of finite degree. Then, 
  \[
    \norm{\left[\left[\BB_{\T}, \BB_{1\Oc\T}\right], \BB_{2\Oc\T}\right]} \leq {C_N}
    ( 1 + \abs{\T} )^{-N}
  \]
  The same estimate holds if we replace one or more operators by their adjoints.
  \proof[Proof] By a smooth decomposition of the wave packet 
  \(\tilde f = \tilde f_{1^c} + \tilde f_{2^c}\), such
  that the resulting commutators~\([\BB^{k^c}_{\T}, \BB_{k\Oc\T}]\) are both
  rapidly decreasing in norm, the result follows directly from \Cref{lem:estNoneq}
  and the Jacobi identity. \qed
\end{Cor}

\noindent The results of this section seem to be somewhat similar in spirit to
Theorem 2 ($ii$\hspace{1pt}) of
\cite{Hdg13}, although their role in our verification of convergence of scattering
states by discretized time sequences is quite different. A similar result can
be found in \cite{MD13}.

\section{Large space-like translations and clustering}
\label{sec:cluM}
In this section we prove the following clustering property for the
operator \tfamilies~\(\AA_{k\Oc\T}\), 
\[
\lim_{\T\rightarrow\infty} E_\Omega^\perp [\AA_{1\Oc\T}^*, \AA_{2\Oc\T}]\Omega = 0,
 \label{eq:clusterI}\numberthis
\]
with \(E_\Omega := \left|\Omega \right\rangle\!\left\langle \Omega \right|\),
\(E_\Omega^\perp := \Id - E_\Omega\), and where in contrast to \Cref{sec:commut}
no restrictions are imposed on velocity supports.
We will require that the scaling \(\mu > 0\) has been chosen sufficiently small
(depending on the Reeh-Schlieder degrees). 
Combined with the single-particle convergence established in \Cref{prop:elemProp}
\itref{it:singleP}, relation \eqref{eq:clusterI} implies that also the limit of
\([\AA_{1\Oc\T}^*,\AA_{2\Oc\T}]\Omega\) exists
and is proportional to the vacuum. Similarly as in \Cref{sec:commut} we will
admit some relative translations of the two operators in~\eqref{eq:clusterI}, so
that the results can be carried over to the corresponding
expressions involving the operators~\(\BB_{k\Oc\T}\) in \Cref{sec:cluConseq}.
These estimates will play a key role for our proof of convergence of scattering
states. 

Our treatment is chiefly inspired by Section~3 of \cite{Dy05} and corresponding
earlier results of Buchholz~\cite{Bu77}. We rely similarly on space-like
decay of matrix elements of local operators, as established by the well-known
Araki-Hepp-Ruelle Theorem. For smooth operators~\(B \in \Alg_0(\Reg)\) a variant
of this decay estimate may be conveniently expressed in terms of the
norm~\(\NAHR{B} := \norm{B} + \norm{\partial_0 B}\).

\begin{Thm}[Araki-Hepp-Ruelle \cite{AHR62}]
  \label{conj:AHR}
  Let \(\Aa_k \in \Alg_0(\DoubleCone_{R_{k}})\), \(k=1,2\).
  Then for any \(\abs{\vec x} \geq 2(R_1+R_2)\), we have
  \begin{align*}
    \abs{
    \left\langle \Omega,
      \Aa_1 U(\vec x) E_\Omega^\perp \Aa_2\Omega
      \right\rangle
    } &\leq \frac{\CAHR(R_1+R_2)^\s}{\abs{\vec x}^{\sMOne}}
      \NAHR{\Aa_1} \NAHR{\Aa_2}.
      \label{eq:cluster2}\numberthis
  \end{align*}
  The constant \(\CAHR\) is universal, but we note that estimate
  \eqref{eq:cluster2} with its quadratic decay is specific to theories on
  physical Minkowski space-time~\(\RealNum^4\).
\end{Thm}

To establish the clustering estimate~\eqref{eq:clusterI} we will have to assume that \(\Aa_\beta \in
\Alg_0(\Reg)\) for small enough \(\beta>0\) and that \(\NAHR{\Aa_\beta}\) is not
growing too fast. Both assumptions follow from the uniform differentiability
property discussed in \Cref{rem:unifDiff}.
Further we will make use of the velocity support estimate of \Cref{lem:velo}
supplemented by well-known globally valid norm estimates for Klein-Gordon
solutions, which we collect in \Cref{lem:decayWave}.

\begin{Prop}
  \label{lem:decayWave}
  Let \(f\) be a regular solution of the Klein-Gordon equation with mass \(m
  \geq 0\) and set \(f_\T(\vec x):= f(\T,\vec x)\).
  Then for any \(p \geq 1\) and \(0 < \epsilon < 1\) the following estimates hold.
  \begin{enumerate}[(i)]
    \item 
      \( \abs{f(t, \vec x)} \leq C_{N} \epsilon^{-N}(1+\abs{t} + \abs{\vec
      x})^{-N}\,\)
      for \(\abs{\vec x} \geq (1+\epsilon)\abs{t}\) and any \(N \in
      \NaturalNum\).
    \item[(i$_0$)]
      If \(m = 0\), then (i) holds also for any 
       \(\abs{\vec x} \leq (1-\epsilon)\abs{t}\).
   \item For \(m > 0\), we have \(\norm{f_\T}_\infty \leq C
     (1+\abs{\T})^{-\s/2}\)  everywhere.
    \item[(ii$_0$)] \(\norm{f_\T}_\infty \leq C (1+\abs{\T})^{-1}\)  everywhere.
    \item For \(m > 0\), \(\norm{f_\T}_{p}^p \leq C_{p}
      (1+\abs{\T}^{\frac \s 2 \cdot (2-p)})\).
    \item[(iii$_0$)] If \(m=0\), then 
      \(\norm{f_\T}_{p}^p \leq C_{\epsilon,p} (1+\abs{\T}^{
    2-p +\epsilon }) \)
      for any \(\epsilon > 0\).
    \item \(\norm{f_\T}_{2} 
        = (2\pi)^{-\s/2} \, \normm{\tilde f}_{2}\) is constant (even if \(m
      = 0\)).
  \end{enumerate}
  All appearing constants depend on the wave packet of \(f\) and norms are 
  taken in \(\LSpace^p(\RealNum^\s)\).
  \proof
  \itr{i} and \itr{i_0} can be established as consequences of the velocity support
  estimate of \Cref{cor:asympt}. Note that for
  \(m=0\) we assumed \(\vec
  0 \not \in \support \tilde f\). The global estimates
  \itr{ii} and \itr{ii_0} are proven e.g.\ in \cite{RS3}, Theorems  XI.17 and
  XI.18.
  \itr{iii} and \itr{iii_0} with \(\epsilon = 1\) follow by decomposing the
  integration according to the regions of validity of the respective versions of \itr{i},
  \itr{ii}, i.e.\ for \(m = 0\) we may take \(I_\T := \{ \abs{\abs{\vec
  x}-\abs{\T}} \leq d \abs{\T}\}\) and its complement.
  The present result for \itr{iii_0}  with \(0<\epsilon<1\) follows by setting 
  \(d = d(\T) = \abs{\T}^{-\nu} \) for any \(0<\nu<1\) with \(\nu :=
  1-\epsilon\) and by making use of \Cref{lem:velo}. Finally, \itr{iv} is a
  consequence of the Plancherel identity.  \qed
\end{Prop}

\begin{Lem}
  \label{lem:cluEstLSZ}
  Let the creation-operator approximants \(\AA_{k\Oc\T}\) be defined in terms of
  operator \tfamilies\ \(\Aa_{1\Oc\beta}\) and \(\Aa_{2\Oc\beta}\) which are localized in
  the standard double cone~\(\DoubleCone_R\) (\(R > 0\)).
  For any \(x_1, x_2 \in \RealNum^{\sPOne}\), we have 
  \[
    \norm{E_\Omega^\perp
    [ \AA_{1\Oc\T}(x_1)^*, \AA_{2\Oc\T}(x_2) ] \Omega}^2
    \leq \frac{C(R+\abs{x_2-x_1}_c)^{9} }{\abs{\T}^{\kappa}} \cdot  
    \NAHR{\Aa_{1\Oc\beta(\T)}}^2
    \NAHR{\Aa_{2\Oc\beta(\T)}}^2,
    \label{eq:clusterLSZ} \numberthis
  \]
  where \(\abs{x}_c := \abss{x^0} + \abs{\vec x}\).
  Here \(\kappa = 3/2\) in the case of \(m > 0\) and for \(m=0\) we can choose
  \(\kappa = 1 - \epsilon\) for any \(\epsilon >0\) with \(C\) depending on
  \(\epsilon\) and the wave packets \(\tilde f_k\).
 \proof By translation invariance, it is sufficient to establish
  the estimate for the relative translation by \(x := x_2-x_1\).
  We may express the norm square as a vacuum expectation value
  by writing
  \begin{align*}
    \fl[2em]\norm{E_\Omega^\perp [ \AA_{1\Oc\T}^*, \AA_{2\Oc\T}(x) ] \Omega}^2
    =\abs{\left\langle \Omega, [\AA_{2\Oc\T}(x)^*, \AA_{1\Oc\T}]  E_\Omega^\perp
      [ \AA_{1\Oc\T}^*, \AA_{2\Oc\T}(x) ] \Omega\right\rangle}
    \\& =
    \abs{ \int \DInt[\s] x_1 \ldots \DInt[\s]x_4  \;
      {f_{2\Oc\T}^*(\vec x_1)}
      {f_{1\Oc\T}(\vec x_2)}
      {f_{1\Oc\T}^*(\vec x_3)}
      {f_{2\Oc\T}(\vec x_4)}
     K(\T,x, \vec x_1, \ldots, \vec x_4)
   }
    \\& \leq
    \int \DInt[\s] x_1 \ldots \DInt[\s]x_4 
    \abs{f_{2\Oc\T}( \vec x_1)}
    \abs{f_{1\Oc\T}( \vec x_2)}
    \abs{f_{1\Oc\T}( \vec x_3)}
    \abs{f_{2\Oc\T}( \vec x_4)}
    \cdot \abs{K(\T,x, \vec x_1, \ldots, \vec x_4)},
    \label{eq:cluInt}\numberthis
  \end{align*}
  where due to time-translation invariance the matrix element \(K\) can be written as
  \begin{align*}
    K &:= 
    \left\langle \Omega, 
       [\Aa_{2\Oc\beta,x}(\vec x_1)^*, \Aa_{1\Oc\beta}(\vec x_2)]  
       E_\Omega^\perp
    [ \Aa_{1\Oc\beta}( \vec x_3)^*, \Aa_{2\Oc\beta,x}(\vec x_4) ] 
    \Omega\right\rangle.
  \end{align*}
  For compact notation, we introduced the abbreviation \(\Aa_{2\Oc\beta,x} :=
  \alpha_x(\Aa_{2\Oc\beta})\) and we suppressed the
  \(\T\)-dependence of \(\beta = \beta(\T)\).

  Now we can estimate \(K\) by combining its support properties resulting from
  locality~\eqref{eq:HK2} with the space-like decay estimates from
  \Cref{conj:AHR} in a manner which seems to be originally due to
  Buchholz~\cite{Bu77}.
  More precisely, by covariance, isotony and the geometry of double cones, the
  standard double cone \(\DoubleCone_{R_2+\abs{x}_c}\) provides a localization
  region for the translated operator \tfamily\ \(\Aa_{2\Oc\beta, x}\).  Therefore
  \(K\) can only be non-zero if 
  \[
    \begin{aligned}
      \abs{\vec x_1 - \vec x_2} &\leq R_1+R_2+\abs{x}_c\\
      \abs{\vec x_3 - \vec x_4} &\leq R_1+R_2+\abs{x}_c
    \end{aligned}
    \label{eq:commBall}\numberthis
  \]
  are both satisfied. This (for fixed \(x\)) finite restriction on the
  relative differences \(\vec x_1-\vec x_2\) and \(\vec x_3-\vec x_4\) now
  allows for successfully estimating the integrand of \eqref{eq:cluInt} for
  large enough relative distance \(\vec x_2-\vec x_3\) ``across'' \(E_\Omega^\perp\) by
  means of
  \Cref{conj:AHR}.

  Restricting the integral \eqref{eq:cluInt} to the region subject to the
  constraints \eqref{eq:commBall}, which we shall denote by \(M \subset
  (\RealNum^{\s})^{ 4}\),
  we find that for points \((\vec x_1, \ldots, \vec x_4) \in M\),
  the two appearing commutators can be localized in suitably translated double
  cones, whose radii can be simultaneously bounded from above
  by~\(R':=2(R_1+R_2)+\abs{x}_c\), i.e.
  \begin{align*}
    C_1:= [\Aa_{2\Oc\beta,x}(\vec x_1)^*, \Aa_{1\Oc\beta}(\vec x_2)]  
         & \in \Alg(\Reg_{\vec x_2}),
         \quad \Reg_{\vec x_2}:=
         \DoubleCone_{R'} + (0, \vec x_2),
         \\
    C_2:=   [\Aa_{1\Oc\beta}(\vec x_3)^*, \Aa_{2\Oc\beta,x}(\vec x_4)]  
       & \in \Alg(\Reg_{\vec x_3}),
       \quad \Reg_{\vec x_3}:=
       \DoubleCone_{R'} + (0, \vec x_3).
  \end{align*}
  Note that \(C_1\) and \(C_2\) are both differentiable by the product rule, as
  a consequence of the assumed  differentiability of the
  \tfamilies~\(\Aa_{k\Oc\beta}\).  
  To apply \Cref{conj:AHR} we subdivide \(M\) into the region 
  \( M_1 := \{(\vec x_1,\ldots,\vec x_4)\in M: \abs{\vec x_2 - \vec x_3} > 2R'\}\)
  and its complement \(M_2 := M \setminus M_1\) and write
  \begin{align*}
    \norm{E_\Omega^\perp [ \AA_{1\Oc\T}^*, \AA_{2\Oc\T}(x) ] \Omega}^2
    &\leq I_{M_1} + I_{M_2},
  \end{align*}
  where \(I_{M_k}\) denotes the integration part of \eqref{eq:cluInt} over the
  subregion~\(M_k\).
  On \(M_1\) we have by \Cref{conj:AHR},
  \[
    \abs{K} \leq \frac{\CAHR (2R')^\s}{\abs{\vec x_2 - \vec x_3}^{\sMOne}} C_\Aa,
    \quad C_\Aa:= \NAHR{C_1} \NAHR{C_2} \leq
    4\NAHR{\Aa_{1\beta}}^2\NAHR{\Aa_{2\beta}}^2.
  \]
  Also note that trivially \(\abs{K} \leq C_\Aa\) holds everywhere.
  Here we made use of 
  \begin{align*}
    \NAHR{C_2} 
    &\leq \norm{[\Aa_1^*, \Aa_2]} + \norm{[\partial_0 \Aa_1^*, \Aa_2]}
    + \norm{[ \Aa_1^*, \partial_0\Aa_2]}
    \\&\leq 2(\norm{\Aa_1^*} \norm{\Aa_2} + \norm{\partial_0 \Aa_1^*}\norm{\Aa_2}
      + \norm{\Aa_1^*} \norm{\partial_0 \Aa_2})
      \\&\leq 2\NAHR{\Aa_1^*} \NAHR{\Aa_2} = 2\NAHR{\Aa_1}\NAHR{\Aa_2},
  \end{align*}
  and similarly for \(\NAHR{C_1}\), where we suppressed dependencies on
  \(\beta\), \(x\) and \(\vec x_k\).  This allows us to estimate
  \begin{align*}
    I_{M_1} &=\int\limits_{M_1} \DInt[\s] x_1 \ldots \DInt[\s]x_4 
    \abs{f_{2\Oc\T}( \vec x_1)}
    \abs{f_{1\Oc\T}( \vec x_2)}
    \abs{f_{1\Oc\T}( \vec x_3)}
    \abs{f_{2\Oc\T}( \vec x_4)}
      \cdot \abs{K(\T, x, \vec x_1, \ldots, \vec x_4)}
    \\&\leq\int\limits_{M_1} \DInt[\s] x_1 \ldots \DInt[\s]x_4 
    \abs{f_{2\Oc\T}( \vec x_1)}
    \abs{f_{1\Oc\T}( \vec x_2)}
    \abs{f_{1\Oc\T}( \vec x_3)}
    \abs{f_{2\Oc\T}( \vec x_4)}
    \frac{\CAHR (2R')^\s}{\abs{\vec x_2 - \vec x_3}^{\sMOne}} C_\Aa
    \\&=\CAHR C_\Aa (2R')^\s 
    \int\limits_{\mathclap{\abs{\vec x_2 - \vec x_3}>2R'} } 
    \DInt[\s] x_2  \DInt[\s]x_3
  \frac{ \abs{f_{1\Oc\T}( \vec x_2)} \abs{f_{1\Oc\T}( \vec x_3)}}
    {\abs{\vec x_2 - \vec x_3}^{\sMOne}} 
    \int\limits_{\mathclap{\abs{\vec x_1 - \vec x_2}<R'} } 
    \DInt[\s] x_1 \abs{f_{2\Oc\T}( \vec x_1)}
    \int\limits_{\mathclap{\abs{\vec x_3 - \vec x_4}<R'} } 
    \DInt[\s]x_4 \abs{f_{2\Oc\T}( \vec x_4)}
    \\&\leq
    (2R')^{9} \norm{f_{2\Oc\T}}_\infty^2 \cdot \CAHR  C_\Aa   \cdot
    \int\limits_{\mathclap{\abs{\vec x_2 - \vec x_3} > 2R'}} 
    \DInt[\s] x_2 \DInt[\s]x_3 \;
    \abs{f_{1\Oc\T}( \vec x_2)}
    \abs{f_{1\Oc\T}( \vec x_3)}
      \frac{1}{\abs{\vec x_2 - \vec x_3}^{\sMOne}}.
    \label{eq:subintEst}\numberthis
  \end{align*}
  Here and below, all appearing \(p\)-norms (\(1\leq p \leq \infty\))
  are on \(\LSpace^p(\RealNum^\s)\)-spaces associated to spatial smearing
  functions.
  We proceed by first estimating the \(\DInt[\s]x_3\)
  subintegral for fixed \(\vec x_2\) using Cauchy-Schwarz (all integrals below over
    \(\{ \vec x_3 \in \RealNum^{\s}: \abs{\vec x_2 - \vec x_3} > 2R'\}\))
  \begin{align*}
    \int
    \DInt[\s]x_3 \; \frac{ \abs{f_{1\Oc\T}( \vec x_3)} }{\abs{\vec x_2 - \vec
 x_3}^{\sMOne}}
   & \leq 
 \norm{f_{1\Oc\T}}_{2}
 \cdot
 \left(\int \frac{\DInt[\s]x_3 }{\abs{\vec x_2 - \vec x_3}^{4}}\right)^{1/2}
 \leq C_{R^{-1}} \normm{ f_{1\T}}_2 .
  \end{align*}
  Here both terms are uniformly bounded in \(\T\), by the Plancherel identity or
  explicit computation\footnote{Performing the second integral in spherical
    coordinates around~\(\vec x_2\) leads to the radial integration beginning
    at \(2R' > R > 0\), which can be
  estimated uniformly in \(\abs{x}_c\) by a finite constant \(C_{R^{-1}}\).},
  respectively.
  
Plugging this into the remaining \(\DInt[\s]x_2\)-integration in
\eqref{eq:subintEst}, we have now shown that 
  \[ 
    I_{M_1} \leq \CAHR C_\Aa C_{R^{-1}}  (2R')^{9} \norm{f_{2\Oc\T}}_\infty^2
                \norm{f_{1\Oc\T}}_2 \norm{f_{1\Oc\T}}_1.
  \]
  On \(M_2\) we estimate similarly using \(\abs{K}\leq C_\Aa\),
  \begin{align*}
    I_{M_2} &
      \leq\int\limits_{M_2} \DInt[\s] x_1 \ldots \DInt[\s]x_4 
    \abs{f_{2\Oc\T}( \vec x_1)}
    \abs{f_{1\Oc\T}( \vec x_2)}
    \abs{f_{1\Oc\T}( \vec x_3)}
    \abs{f_{2\Oc\T}( \vec x_4)}
     C_\Aa
    \\&\leq
     C_\Aa (2R')^{9} \norm{f_{2\Oc\T}}_\infty^2  \norm{f_{1\Oc\T}}_\infty 
     \norm{f_{1\Oc\T}}_1.
  \end{align*}
  The result now follows from \Cref{lem:decayWave}.\qed
\end{Lem}

\section{Consequences of the clustering estimate}
\label{sec:cluConseq}
With the clustering estimate for the operators~\(\AA_{k\T}\) from
\Cref{lem:cluEstLSZ} at hand, it is straightforward to prove clustering of the
creation operators \(\BB_{k\T}\) for Reeh-Schlieder \tfamilies\ of finite degree.

\begin{Prop}
  \label{lem:cluMD}
  For uniformly differentiable Reeh-Schlieder \tfamilies~\(\Aa_{1\Oc\beta}\),
  \(\Aa_{2\Oc\beta}\), and regular Klein-Gordon
  solutions~\(f_1\), \(f_2\) of mass~\(m \geq 0\), we have 
  \begin{align*}
    \norm{E_\Omega^\perp \BB_{1\Oc\T}^* \BB_{2\Oc\T} \Omega}
    &\leq \frac{C}{\abs{\T}^{\kappa/2}} 
    \NAHR{\Aa_{1\Oc\beta(\T)}}
    \NAHR{\Aa_{2\Oc\beta(\T)}}.
  \end{align*}
  Here, \(C\) depends on \(\chi\), localization regions of~\(\Aa_{k\Oc\beta}\),
  wave packets, and \(\kappa\) (see \Cref{lem:cluEstLSZ}). 
  \proof
  As \(\BB_{1\Oc\T}^*\Omega = 0\), we can replace the product
  \(\BB_{1\Oc\T}^*\BB_{2\Oc\T}\)
  acting on the vacuum by the commutator
  \(
    [ \BB_{1\Oc\T}^*, \BB_{2\Oc\T} ] 
  \).
  Making use of \(\BB_{k\T} = \AA_{k\T}(\chi)\), \(\chi \in
  \SchwartzSpace(\RealNum^\sPOne)\), and \Cref{lem:cluEstLSZ}, we obtain
  \begin{align*}
  \fl  \norm{E_\Omega^\perp [ \BB_{1\Oc\T}^*, \BB_{2\Oc\T} ] \Omega}
    \leq \int \DInt[\sOne]x_1 \DInt[\sOne]x_2 \;
    \abs{\chi(x_1)}\abs{\chi(x_2)} \cdot 
    \norm{E_\Omega^\perp [ \AA_{1\Oc\T}^*(x_1), \AA_{2\Oc\T}(x_2) ] \Omega}
  \\
    &\leq \int \DInt[\sOne]x_1 \DInt[\sOne]x_2 \;
    \abs{\chi(x_1)} \abs{\chi(x_2)} 
     \frac{C'(R+\abs{x_1 - x_2}_c)^{9/2}}{\abs{\T}^{\kappa/2}} 
    \NAHR{\Aa_{1\Oc\beta(\T)}}
    \NAHR{\Aa_{2\Oc\beta(\T)}}
  \\ & =
    \frac{C}{\abs{\T}^{\kappa/2}}
    \NAHR{\Aa_{1\Oc\beta(\T)}}
    \NAHR{\Aa_{2\Oc\beta(\T)}}. \qedhere
  \end{align*}
\end{Prop}

For Reeh-Schlieder \tfamilies\ of finite degree \Cref{lem:cluMD} simplifies
further, yielding a constraint for admissible choices of scaling.
In the following \(\gamma\) always denotes the (finite) largest appearing
degree%
, i.e.\ \(\NAHR{\Aa_{k\Oc\beta}}\leq \beta^{-\gamma}\) for small
enough \(\beta > 0\) and all \(k=1,\ldots,n\).
From now on we will also adopt the canonical scaling \(\beta(\T) :=
\abs{\T}^{-\mu}\), \(\mu > 0\).

\begin{Cor} \label{cor:cluster}%
  Let the Reeh-Schlieder \tfamilies~\(\Aa_{1\Oc\beta}\), \(\Aa_{2\Oc\beta}\) have
  finite degrees.
  Under the assumptions of \Cref{lem:cluMD}, there exists a \(C > 0\)
  such that for large enough~\(\T\) we have
  \[
    \norm{E_\Omega^\perp \BB_{1\Oc\T}^* \BB_{2\Oc\T}\Omega} \leq C
    \abs{\T}^{2\gamma\mu-\kappa/2},
  \]
  Consequently for any \(0 < \mu < \frac{\kappa}{4\gamma}\) we obtain
  \[
    \lim_{\T\rightarrow \infty} E_\Omega^\perp \BB_{1\Oc\T}^* \BB_{2\Oc\T}\Omega = 0.
  \]
  \proof Follows immediately from inserting
  \(\NAHR{\Aa_{k\Oc\beta(\T)}} \leq C' \beta(\T)^{-\gamma} = C '\abs{\T}^{\gamma\mu}\)
  into the estimate of \Cref{lem:cluMD}. \qed
\end{Cor}

While \Cref{cor:cluster} will be sufficient to establish the Fock structure of
scattering states in \Cref{sec:Fock}, our proof of convergence relies on an
extension of this result, which is concerned with the case of multiple
creation operators.
The resulting \Cref{lem:multiClust2} combines energy bounds and clustering
estimates in a novel way. It may be considered our main technical result.
\begin{Lem}[multi-operator clustering]
  \label{lem:multiClust2}
  For \(\T_1, \ldots, \T_n \in \RealNum\) denote by 
  \(\abs{\Tmin} > 0\) and \(\abs{\Tmax}\) the minimum and maximum of
  absolute values \(\abs{\T_k}\), (\(1\leq k\leq n\)), respectively. Then
  for large enough \(\Tmin\)
  \[
    \norm{  E_\Omega^\perp \left( \prod_{k=1}^n \BB^*_{k\Oc\T_k} \BB_{k\Oc\T_k} \right)\!\Omega} 
    \leq C  \abs{\Tmax}^{2n\gamma\mu} \cdot \abs{\Tmin}^{-\kappa/2}.
    \label{eq:moce} \numberthis
  \]
  The constant~\(C\) is independent of the \(\T_k\), but depends on the number of
  pairs~\(n\),  wave packets, Reeh-Schlieder \tfamilies, and the smearing
  function \(\chi\).
 \proof We will show by induction that 
  \[
    \norm{  E_\Omega^\perp 
    \left( \prod_{k=1}^n \BB^*_{k\Oc\T_k} \BB_{k\Oc\T_k} \right)\!\Omega}
    \leq C \sum_{j=1}^n \left(
      \prod_{\substack{k=1}}^{j-1} \norm{E(\Delta)\BB^*_{k\Oc\T_k}\BB_{k\Oc\T_k}E(\Delta)}
    \right)
    \norm{E_\Omega^\perp \BB_{j\Oc\T_j}^* \BB_{j\Oc\T_j}\Omega},
    \label{eq:multiest}\numberthis
  \]
  where \(\Delta\subset\RealNum^{\sPOne}\) is a large enough compact set depending
  on \(\support \hat \chi\) and the number of pairs~\(n \in \NaturalNum\).
  From this we obtain by applying 2-operator clustering (\Cref{cor:cluster}), the
  energy bound of \Cref{lem:bu}, and the finite-degree Reeh-Schlieder estimates
  that for large enough \(\abs{\Tmin}\),
  \eqref{eq:moce} holds as claimed. For \(n=1\), statement \eqref{eq:multiest}
  has been established in \Cref{cor:cluster}.
  Assuming that \eqref{eq:multiest} holds for \(n-1\)~pairs, we write
  \begin{align*}
    \norm{E_\Omega^\perp \left( \prod_{k=1}^n \BB^*_{k\Oc\T_k} \BB_{k\Oc\T_k} \right)\!\Omega}
    &
    = \norm{  E_\Omega^\perp \left( \prod_{k=1}^{n-1} \BB^*_{k\Oc\T_k} \BB_{k\Oc\T_k} \right) \!
 (E_\Omega + E_\Omega^\perp) \BB^*_{n\Oc\T_n} \BB_{n\Oc\T_n} \Omega}
    \\
    & \leq \norm{  E_\Omega^\perp \left( \prod_{k=1}^{n-1} \BB^*_{k\Oc\T_k} \BB_{k\Oc\T_k} \right) \!
        E_\Omega \BB^*_{n\Oc\T_n} \BB_{n\Oc\T_n} \Omega}
    \\ & \qquad\qquad
     + \norm{  E_\Omega^\perp \left( \prod_{k=1}^{n-1} \BB^*_{k\Oc\T_k} \BB_{k\Oc\T_k} \right) \!
        E_\Omega^\perp \BB^*_{n\Oc\T_n} \BB_{n\Oc\T_n} \Omega}.
        \label{eq:cluStep}\numberthis
  \end{align*}
  Now we would like to estimate the second term by 2-operator clustering
  (\Cref{cor:cluster}). Regarding the applicability of energy bounds
  from \Cref{lem:bu}, we note that by \Cref{prop:elemProp} \itref{it:emTrafo},
  \(\BB_{n\Oc\T_n}^*\BB_{n\Oc\T_n}\) has compact energy-momentum transfer
  \(\Delta' := \support \hat \chi - \support \hat \chi\).
  Therefore we can insert an energy-momentum projection onto \(\Delta'\) (which
  commutes with \(E_\Omega^\perp\)) and estimate \begin{align*}
    \fl[8em] 
    \norm{E_\Omega^\perp \left( \prod_{k=1}^{n-1} \BB^*_{k\Oc\T_k} \BB_{k\Oc\T_k} \right)\!   
    E(\Delta')  E_\Omega^\perp \BB^*_{n\Oc\T_n} \BB_{n\Oc\T_n} \Omega}
    \\
    &\leq
      \norm{E_\Omega^\perp \left( \prod_{k=1}^{n-1} \BB^*_{k\Oc\T_k} \BB_{k\Oc\T_k} \right)\!
      E(\Delta')}
      \cdot \norm{E_\Omega^\perp \BB^*_{n\Oc\T_n} \BB_{n\Oc\T_n} \Omega} 
    \\ &
      \leq 
      \left(
      \prod_{k=1}^{n-1} \norm{E(\Delta)  \BB^*_{k\Oc\T_k} \BB_{k\Oc\T_k}
    E(\Delta)}\right)
\cdot 
\norm{E_\Omega^\perp \BB^*_{n\Oc\T_n} \BB_{n\Oc\T_n} \Omega},
  \end{align*}
  where we have chosen the compact set \(\Delta \subset\RealNum^{\sPOne}\) large
  enough (depending on \(n\)) to contain the sum of the energy-momentum
  transfers differences~\(\Delta'\) of all creation-annihilation operator pairs.
  Similarly we estimate the first term in \eqref{eq:cluStep} by 
  making use of the one-dimensional nature of the projection~\(E_\Omega\), 
  and the induction assumption,
  \begin{align*}
    \fl
    \norm{  E_\Omega^\perp \left( \prod_{k=1}^{n-1} \BB^*_{k\Oc\T_k} \BB_{k\Oc\T_k} \right)\!
 E_\Omega \BB^*_{n\Oc\T_n} \BB_{n\Oc\T_n} \Omega}
 =
      \norm{  \BB_{n\Oc\T_n}\Omega}^2
      \cdot \norm{E_\Omega^\perp \left( \prod_{k=1}^{n-1} \BB^*_{k\Oc\T_k} \BB_{k\Oc\T_k} \right)\!\Omega}
    \\&
      \leq C
      \cdot \sum_{j=1}^{n-1} \norm{  E_\Omega^\perp \BB^*_{j\Oc\T_j} \BB_{j\Oc\T_j} \Omega}
      \cdot \left( \prod_{\substack{k=1}}^{j-1}
      \norm{E(\Delta)\BB^*_{k\Oc\T_k}\BB_{k\Oc\T_k}E(\Delta)} \right).
  \end{align*}
  Here we also made use of the fact that \(\norm{  \BB_{n\Oc\T_n}\Omega} \leq
    C\) is uniformly bounded in \(\T_n\) by convergence to the corresponding
    single-particle state (see \Cref{prop:elemProp} \itref{it:singleP}).
  Taken together, these two estimates complete the induction step.
  \qed
\end{Lem}

A useful consequence of multi-operator clustering, which will be important
for us later, is the boundedness of scattering-state approximants, i.e.\ vectors
resulting from iterated application of creation-operators to the vacuum.
In fact, a similar result was used by Buchholz for the collision
theory of massless bosons~\cite{Bu77}. While the proofs of Buchholz' results
can be simplified\footnote{See e.g.\ \cite{AD15}.} using methods from harmonic
analysis \cite{Bu90}, our construction is based on operator
\tfamilies~\(\Aa_\beta\) with diverging norms in the limit~\(\beta\rightarrow
0\). This norm growth will be inherited by energy bounds for creation operators,
if they are derived by means of \Cref{lem:bu}. In the vacuum
sector of a local theory however, we can establish uniform estimates on
scattering-state approximants by relying on the good behaviour of
\(\Aa_\beta\Omega\) via the previously established clustering properties,
similarly as in \cite{Bu77}.
\begin{Cor}
  \label{cor:clusterBound}
  Assume disjoint velocity supports.
  For any scaling \(0 < \mu < \tfrac\kappa{4\gamma(n-1)}\), there exists a \(C >
    0\), such that for all sufficiently large \(\T \in \RealNum\) and all \(\T_k\) from the
    corresponding interval spanned by \(\T\) and \( \T+\rho\T\), 
  \[ 
    \norm{\BB_{1\Oc\T_1} \ldots \BB_{n\Oc\T_n} \Omega} \leq C,
  \]
  with \(\rho\) as in \Cref{lem:estNoneq}
  (for \(n=1\), any \(\mu \in (0,\infty)\) is admissible).
 \proof The proof is by induction on the number of particles~\(n\).
  For \(n=1\), the claim follows by convergence to the corresponding
  single-particle state as proven in \Cref{prop:elemProp} \itref{it:singleP}.
  For the general case it will be sufficient to establish the claim for large
  enough \(\abs{\T}\), as can be seen from the simple polynomial estimate of
  \Cref{prop:elemProp}. Let us now assume that the statement holds for \(n\)
  particles.
  For simplicity we set \(\BB_k:=\BB_{k\Oc \T_k}\) and write
  \def\TT#1{\T_{#1}}
  \def\TT#1{}
  \begin{align*}
    \norm{\BB_{1\Oc\TT 1} \ldots \BB_{n+1\NOc\TT {n+1}} \Omega}^2
    &= 
    \left\langle \Omega, \BB_{n+1\NOc\TT {n+1}}^* \ldots \BB_{1\Oc\TT 1}^*
       \BB_{1\Oc\TT 1} \ldots \BB_{n+1\NOc\TT {n+1}}\Omega\right\rangle
    \\&=
    \left\langle \Omega, \BB_{n+1\NOc\TT {n+1}}^* \BB_{n+1\NOc\TT {n+1}} 
       \BB_{n\Oc\TT n}^*\ldots \BB_{1\Oc\TT 1}^* \BB_{1\Oc\TT 1} \ldots \BB_{n\Oc\TT n} \Omega\right\rangle 
    \\&\qquad
    +  \left\langle \Omega, \BB_{n+1\NOc\TT {n+1}}^* \left[ \BB_{n\Oc\TT n}^*\ldots \BB_{1\Oc\TT 1}^*
        \BB_{1\Oc\TT 1} \ldots \BB_{n\Oc\TT n}, \BB_{n+1\NOc\TT {n+1}}\right]
      \Omega\right\rangle,
  \end{align*}
  where the absolute value of the second term is bounded, as it vanishes for
  \(\abs{\T}\rightarrow \infty\) for any choice of scaling by the rapid decay of
  commutators (\Cref{lem:estNoneq}). This decay can compensate the norm growth of the
  creation-operator approximants, which is at most polynomial --- even when using the naive
  estimate of \Cref{prop:elemProp}. 

  Therefore it is sufficient to establish boundedness of the matrix element
  \begin{align*}
    \fl
    \left\langle \Omega, \BB_{n+1\NOc\TT {n+1}}^* \BB_{n+1\NOc\TT {n+1}}
      \BB_{n\Oc\TT n}^*\ldots
      \BB_{1\Oc\TT 1}^* \BB_{1\Oc\TT 1} \ldots \BB_{n\Oc\TT n} \Omega\right\rangle  
      \\&=
      \left\langle \Omega, \BB_{n+1\NOc\TT {n+1}}^* \BB_{n+1\NOc\TT {n+1}} (E_\Omega +
        E_\Omega^\perp) \BB_{n\Oc\TT n}^*\ldots \BB_{1\Oc\TT 1}^* \BB_{1\Oc\TT 1} \ldots \BB_{n\Oc\TT n} \Omega\right\rangle  
        \\&= \norm{\BB_{n+1\NOc\TT {n+1}}\Omega}^2 \norm{\BB_{1\Oc\TT 1} \ldots \BB_{n\NOc\TT n} \Omega}^2
      + 
      \left\langle \Omega, \BB_{n+1\NOc\TT {n+1}}^* \BB_{n+1\NOc\TT {n+1}}  E_\Omega^\perp
        \BB_{n\Oc\TT n}^*\ldots \BB_{1\Oc\TT 1}^* \BB_{1\Oc\TT 1} \ldots \BB_{n\Oc\TT n} \Omega\right\rangle.  
        \label{eq:boundind} \numberthis
  \end{align*}
  The first term of \eqref{eq:boundind} provides the dominant contribution in the limit
\(\abs{\T}\rightarrow\infty\): its two factors are bounded by the induction
  assumption and the one-particle case. The second term can be written as the
  sum of 
  \(
    \left\langle \Omega, \BB_{n+1\NOc\TT {n+1}}^* \BB_{n+1\NOc\TT {n+1}}  
        E_\Omega^\perp \BB_{n\Oc\TT n}^* \BB_{n\Oc\TT n} \ldots \BB_{1\Oc\TT 1}^* \BB_{1\Oc\TT 1}
     \Omega\right\rangle 
  \) 
  and further matrix elements involving at least one commutator of operators
  involving disjoint velocity supports.
  As before, the latter are rapidly decreasing by \Cref{lem:estNoneq}.
  We can conclude the proof by applying the Cauchy-Schwarz inequality to the
  remaining term 
  \begin{align*}
    \abs{
      \left\langle \Omega, \BB_{n+1\NOc\TT {n+1}}^* \BB_{n+1\NOc\TT {n+1}}  
        E_\Omega^\perp \BB_{n\Oc\TT n}^* \BB_{n\Oc\TT n} \ldots \BB_{1\Oc\TT 1}^* \BB_{1\Oc\TT 1}
     \Omega\right\rangle }
     &\leq \norm{E_\Omega^\perp \BB_{n+1\NOc\TT {n+1}}^* \BB_{n+1\NOc\TT {n+1}}\Omega} 
     \cdot \norm{E_\Omega^\perp \BB_{n\Oc\TT n}^*\BB_{n\Oc\TT n} \ldots \BB_{1\Oc\TT 1}^* \BB_{1\Oc\TT 1}  \Omega}
  \end{align*}
where both factors vanish in the limit \(\abs{\T} \rightarrow \infty\) for any sufficiently
  small choice of scaling~\(\mu\) by \Cref{lem:multiClust2}. \qed
\end{Cor}
\section{Convergence of scattering state approximants}
\label{sec:conv}
For this section we adopt the standing assumptions that \(\Aa_{1\Oc\beta},
\ldots, \Aa_{n\beta}\) are Reeh-Schlieder \tfamilies\ of finite degree and we take
\(f_1, \ldots, f_n\) to be regular positive-energy Klein-Gordon solutions of the
corresponding mass with pairwise disjoint velocity supports.  

\begin{Thm}
  \label{thm:conv}
  Let the Reeh-Schlieder \tfamilies~\(\Aa_{1\Oc\beta}, \ldots,
  \Aa_{n\beta}\) have degrees less than some finite value~\(\gamma > 0\) and 
  take a scaling exponent \(\mu \in (0, \tfrac{\kappa}{ 4(n-1)\gamma })\)
  (\(\kappa\) as in \Cref{lem:cluEstLSZ}).
  \begin{enumerate}[(i)]
    \item The \tfamily\
      \( 
        \Psi_\T := \BB_{1\Oc\T} \ldots \BB_{n\Oc\T} \Omega
      \)
      is convergent in norm as \(\T\rightarrow \pm \infty\).
    \item
      The limit is independent of the choice of \(\mu\), \(\Aa_{k\Oc\beta}\) 
      and \(f_k\) within the specified restrictions, as long as the associated operators
      \(\BB_{k\Oc\T}'\) create on the vacuum the same family of single-particle
      states~\(\Psi_{k}^{(1)} = \lim_{\T\rightarrow \infty} \BB_{k\Oc\T}\Omega\).
  \end{enumerate}
\end{Thm}

Avoiding differentiability assumptions on~\(\Aa_{k\Oc\beta}\) with respect to
the parameter~\(\beta\), we will proceed by a discrete variant of Cook's method,
thereby reducing the convergence of the scattering state
approximants~\( \Psi_\T \)
to the convergence of single-particle state approximants \(\BB_{k\Oc\T}\Omega\).
Recall that for Reeh-Schlieder \tfamilies~\(\Aa_{k\Oc\beta}\), we have
quantitative control over the convergence of the single-particle problem by
\Cref{prop:elemProp}~\itref{it:singleP}.

The restrictions on the time differences to obtain rapid decay of commutators
in \Cref{lem:estNoneq} suggests to consider the restrictions of \(\Psi_{\T}\) to
sequences
\[
  \T_k = (1+\rho)^k \T_0, \quad \T_0 \not = 0\; \text{arbitrary},
  \label{eq:timeSp}\numberthis
\]
and \( \rho > 0 \) depending on the separation of velocity supports as explained
in \Cref{lem:estNoneq}.

As preparation for proving \Cref{thm:conv} we will first show that 
we can relate the norm of differences \(\Psi_{\T_2} - \Psi_{\T_1}\) to
corresponding single-particle expressions \(\norm{\BB_{k\Oc\T_2}\Omega -
\BB_{k\Oc\T_1}\Omega}\)
at least \emph{``locally''}, i.e.\ if we place sufficient restrictions on the differences
\(\abs{\T_2-\T_1}\). We will give a unified account for proving both parts of
\Cref{thm:conv} by comparing the scattering state approximants associated to two
possibly distinct families of creation operators with comparable velocity
supports.
Thereto let \(\Aa_{k\Oc\beta}\), \(\Aa_{k\Oc\beta}' \in \Alg(\Reg)\) be uniformly
differentiable Reeh-Schlieder \tfamilies\ of finite degree, and choose regular positive-energy Klein-Gordon 
solutions \(f_1, \ldots, f_n\) and \(f_1', \ldots f_n'\) of mass \(m \geq 0\)
such that all pairs with \(j \not = k\) (including mixed pairs \(f_j\),
\(f_k'\)) have disjoint velocity supports.  We denote the corresponding creation
operators by \(\BB_{k\Oc\T}\), \(\BB_{k\Oc\T}'\) and set
\[
  \Psi_\T : = \BB_{1\Oc\T} \ldots \BB_{n\Oc\T} \Omega,
  \quad
  \Psi_\T' : = \BB_{1\Oc\T}' \ldots \BB_{n\Oc\T}' \Omega.
\]

\begin{Rem}[change of scaling] \label{rem:scalVar}
Anticipating also the proof of \Cref{thm:conv} \itr{ii}, we may also allow the
creation operator \tfamilies~\(\BB_{k\Oc\T}\) and \(\BB_{k\Oc\T}'\) to be
defined using distinct choices of scaling \(\beta_k(\T) := \abs{\T}^{-\mu_k}\),
\(\beta_k'(\T) := \abs{\T}^{-\mu_k'}\). On the first reading, this detail can
safely be ignored, but it is easily seen that the statement and proof of
\Cref{lem:localEst} can even be kept invariant under this generalization
if we simply denote the smallest appearing scaling exponent by \(\mu := \min \{ \mu_k,
\mu_k' \; (1\leq k \leq n) \} > 0 \). The required extensions of \Cref{lem:estNoneq},
\Cref{lem:multiClust2}, and \Cref{cor:clusterBound} follow directly by similar
considerations.
\end{Rem}

 \def\dd{\Delta_\T}
\begin{Lem}
  \label{lem:localEst}
  Take \(\rho > 0\) as given in \Cref{lem:estNoneq} (using the smallest value
  suitable for all disjoint pairs of velocity supports), and choose 
  sufficiently small scaling~\(\mu > 0\) (cf.\ \Cref{cor:clusterBound}).
  Then there exist constants \(C_1, C_2 > 0\), such that for sufficiently
  large~\(\abs{\T} > 0\) and any subsequent choice of \(\T_1, \T_2\) from the
interval spanned by  \(\T\) and \( (1+\rho)\T\), we have 
  \[ \norm{\Psi_{\T_2} - \Psi_{\T_1}'} \leq 
        C_1 \sum_{k=1}^n \norm{\BB_{k\Oc\T_2}\Omega - \BB_{k\Oc\T_1}'\Omega}
        + C_2 \abs{\T}^{n\gamma\mu - \kappa/4}.
  \]
 \proof 
  For \(n=1\) the statement is trivial.
  For \(n \geq 2\) we can estimate telescopically
  \begin{align*}
    \norm{ \Psi_{\T_2} - \Psi_{\T_1}' }
    &\leq\sum_{k=1}^n 
    \norm{ \BB_{1\Oc\T_2} \ldots \BB_{k-1\NOc\T_2} (\BB_{k\Oc\T_2} - \BB_{k\Oc\T_1}')
      \BB_{k+1\NOc\T_1}' \ldots \BB_{n\Oc\T_1}'\Omega}.
  \end{align*}
  The claim is obtained if the following estimate can be established for each
  \(1\leq k \leq n\),
  \begin{align*}
    \fl[10em]
    \norm{ \BB_{1\Oc\T_2} \ldots \BB_{k-1\NOc\T_2} (\BB_{k\Oc\T_2} - \BB_{k\Oc\T_1}')
      \BB_{k+1\NOc\T_1}' \ldots \BB_{n\Oc\T_1}'\Omega}^2
    \\
    &\leq
    C_1 \norm{\BB_{k\Oc\T_2}\Omega - \BB_{k\Oc\T_1}'\Omega}^2 + C_2
    \abs{\T}^{2\gamma n\mu - \kappa/2}.
    \label{eq:goal}\numberthis
  \end{align*}
  We will prove this inequality by making use of the rapid decay of
  restricted non-equal time commutators together with the energy bound and
  clustering. Introducing the abbreviation \(\dd \BB_{k}: =  \BB_{k\Oc\T_2} -
  \BB_{k\Oc\T_1}'\), we can write
  \begin{align*} 
    \fl[1em] \norm{\BB_{1\Oc\T_2} \ldots \BB_{k-1\NOc\T_2} (\dd \BB_{k})  
          \BB_{k+1\NOc\T_1}' \ldots \BB_{n\Oc\T_1}'\Omega}^2 
    \\ & = 
        \left\langle  \Omega, \BB_{n\Oc\T_1}'^* \ldots \BB_{k+1\NOc \T_1}'^* (\dd \BB_k)^*
          \BB_{k-1\NOc\T_2}^* \ldots \BB_{1\Oc\T_2}^* \BB_{1\Oc\T_2} \ldots \BB_{k-1\NOc\T_2} (\dd \BB_{k})
          \BB_{k+1\NOc\T_1}' \ldots \BB_{n\Oc\T_1}'\Omega \right\rangle
    \\ & \leq
          \abs{
        \left\langle  \Omega, \BB_{1\Oc\T_2}^* \BB_{1\Oc\T_2} \ldots \BB_{k-1\NOc\T_2}^* \BB_{k-1\NOc\T_2} \cdot
          \BB_{k+1\NOc\T_1}'^*\BB_{k+1\NOc\T_1}' \ldots \BB_{n\Oc\T_1}'^*\BB_{n\Oc\T_1}'  (\dd \BB_k)^* (\dd \BB_{k})
      \Omega \right\rangle} \\ & \qquad \qquad + C_M \abs{\T}^{-M}, 
          \label{eq:ldInterm}\numberthis
  \end{align*}
  and the rapidly decreasing error can be subsumed into the \(C_2\)-term
  of~\eqref{eq:goal}. To obtain \cref{eq:ldInterm}, we made multiple use of the
  non-equal-time commutator estimate\footnote{For the status of
      \Cref{lem:estNoneq} in the context of non-equal scaling, cf.\
    \Cref{rem:scalVar} and Footnote~\ref{foot:commAssump}.}
  of \Cref{lem:estNoneq}, which is
  sufficiently strong for overcompensating to any desired inverse polynomial
  order the asymptotic growth of the elementary estimate~\(\norm{\BB_{k\Oc\T}}
    \leq C_k(1+\abs{\T}^{N + \gamma\mu})\) and similar estimates for adjoints and
    primed operators (see \Cref{prop:elemProp}).

  The remaining term in \eqref{eq:ldInterm} still contains the asymptotically
  dominant contribution, which we will now extract using the clustering
  estimate. Inserting an identity operator~\((E_\Omega + E_\Omega^\perp)\) 
  after \(  (\dd \BB_k^*) (\dd \BB_{k}) \Omega\) and
  making use of subadditivity and decay of commutators yields
  \begin{align*}
    \fl[1em]
    \abs{
    \left\langle \Omega, 
    \BB_{1\Oc\T_2}^* \BB_{1\Oc\T_2} \ldots \BB_{k-1\NOc\T_2}^*  \BB_{k-1\NOc\T_2}  \cdot 
    \BB_{k+1\NOc\T_1}'^*\BB_{k+1\NOc\T_1}' \ldots \BB_{n\Oc\T_1}'^*\BB_{n\Oc\T_1}'
    (\dd \BB_k^*) (\dd \BB_{k}) \Omega \right\rangle}
    \\& \leq
      \normm{ 
E_\Omega^\perp 
      \BB_{n\Oc\T_1}'^*
      \BB_{n\Oc\T_1}'
      \ldots
      \BB_{k+1\NOc \T_1}'^*
      \BB_{k+1\NOc\T_1}' 
      \cdot
      \BB_{k-1\NOc\T_2}^*
      \BB_{k-1\NOc\T_2}
      \ldots
      \BB_{1\Oc\T_2}^*
      \BB_{1\Oc\T_2}
        \Omega 
    }  \cdot
      \normm{(\dd \BB_k^*) (\dd \BB_{k}) \Omega}
    \\&\qquad + 
      \normm{\BB_{1\Oc\T_2} \ldots \BB_{k-1\NOc\T_2}\cdot \BB_{k+1\NOc \T_1}'\ldots
        \BB_{n\Oc\T_1}'\Omega}^2 \cdot \norm{(\dd \BB_k)\Omega}^2 + C_M
        \abs{\T}^{-M}.
  \end{align*}
  Both terms depend on the convergence speed of the single-particle
  problem, although --- anticipating the results of \Cref{sec:Fock} --- we expect
  the second summand to be dominant for large \(\T\): By boundedness of
  scattering state approximants (\Cref{cor:clusterBound}) 
  \[
    \norm{\BB_{1\Oc\T_2} \ldots \BB_{k-1\NOc\T_2}
      \cdot \BB_{k+1\NOc \T_1}'\ldots \BB_{n\Oc\T_1}'\Omega}^2 
    \leq C_1
  \]
  for suitable \(C_1 > 0\).
  It remains to be shown that the first summand has the same asymptotics
  as the \(C_2\)-term of \eqref{eq:goal}. 
  By the clustering result with multiple pairs of creation- and
  annihilation-operator approximants of \Cref{lem:multiClust2}, we obtain that
  \begin{align*} \fl\norm{E_\Omega^\perp \BB_{n\Oc\T_1}'^* \BB_{n\Oc\T_1}'\ldots
      \BB_{k+1\NOc\T_1}'^* \BB_{k+1\NOc\T_1}' \cdot \BB_{k-1\NOc\T_2}^*
      \BB_{k-1\NOc\T_2} \ldots \BB_{1\Oc\T_2}^* \BB_{1\Oc\T_2}\Omega} 
      \leq C_2 \abs{\T}^{2(n-1)\gamma\mu-\kappa/2},
  \end{align*}
  which also made use of the time restriction yielding \(\abs{\T} \leq \abs{\T_{k}} \leq (1+\rho)\abs{\T}\).  The second factor is estimated making use of the energy bound,
  \begin{align*}
  \norm{(\dd \BB_k^*) (\dd \BB_{k}) \Omega}
  &=
  \norm{(\dd \BB_k^*) E(\Delta) (\dd \BB_{k}) \Omega}
  \\&\leq\norm{(\dd \BB_k^*) E(\Delta)} \cdot \norm{ (\dd \BB_{k}) \Omega}
  \\&\leq C_3 \abs{\T_2}^{\gamma\mu } 
  \leq C_3 (1+\rho)^{\gamma\mu} \abs{\T}^{\gamma\mu } 
  =: C_3' \abs{\T}^{\gamma\mu },
  \end{align*}
  where the energy-momentum projection onto the compact set \(\Delta:= \support
  \hat \chi\) can be inserted due to \(\dd \BB_{k} \Omega \in
E(\Delta)\HilbertSpace\)\!. Altogether we obtain~\eqref{eq:goal}, completing
the proof. \qed
\end{Lem}

The convergence of scattering state approximants~\(\Psi_\T\) is now
easily established by iterated application of \Cref{lem:localEst}.

\begin{proof}[Proof of \Cref{thm:conv}] {\em Ad (i)}.
  We estimate by writing a telescopic sum and making use of subadditivity
  of the norm,
  \[
    \norm{\Psi_{\T_L} - \Psi_{\T_0}} 
      \leq \sum_{k=1}^L \norm{\Psi_{\T_k} - \Psi_{\T_{k-1}}}.
  \]
  We have by construction that 
  \(
  \T_k, \T_{k-1}\) are contained in the interval spanned by \( \T_{k-1} \) and
  \( (1+\rho)\T_{k-1} \).
  Thus \Cref{lem:localEst} is applicable with \(\BB_{k\Oc\T} = \BB_{k\Oc\T}'\). 
  Fixing the scaling parameter~\(\mu > 0\) such that \(\delta:= \kappa/4 - n
  \gamma \mu > 0\), all assumptions of \Cref{lem:localEst} are satisfied and we obtain
  \[
    \norm{\Psi_{\T_L} - \Psi_{\T_0}}
    \leq \sum_{k=1}^L \left(
      C_1
      \sum_{j=1}^n \norm{\BB_{j\Oc\T_k}\Omega - \BB_{j\Oc\T_{k-1}}\Omega}
      + C_2 \abs{\T_{k-1}}^{-\delta}\right). 
    \label{eq:localApplied} \numberthis
  \]
  Now, the single-particle convergence property of the Reeh-Schlieder
  \tfamilies\ implies
  \[
    \norm{\BB_{j\Oc\T_k}\Omega - \BB_{j\Oc\T_{k-1}}\Omega}
    \leq
    \norm{\BB_{j\Oc\T_k}\Omega - \Psi_{j}^{(1)}} + \norm{\Psi_{j}^{(1)}- \BB_{j\Oc\T_{k-1}}\Omega}
    \leq C \abs{\T_{k-1}}^{-\mu},
  \]
  where \(\Psi_{j}^{(1)}  = \lim_{\T\rightarrow \pm \infty} \BB_{j\Oc\T}\Omega\).
  Applying this estimate and inserting \( \T_k = (1+\rho)^k \T_0\), we can
  take care of both terms in \eqref{eq:localApplied} by writing
  \[
    \norm{\Psi_{\T_L} - \Psi_{\T_0}}
    \leq C' \sum_{k=1}^L \abs{\T_{k-1}}^{-\mu'} 
    = C' \abs{\T_0}^{-\mu'} \cdot \sum_{k=1}^L (1+\rho)^{-\mu' (k-1)}, 
    \label{eq:gsEst}\numberthis
  \]
  with  \(\mu' := \min(\mu, \delta)\).
  Clearly, the geometric series is convergent for \(L\rightarrow \infty\).
  Independence of the limit from the choice of the sequence \(\T_k\),
  i.e.\ convergence of \(\Psi_\T\) as a function of the continuous
  parameter~\(\T\), may be inferred from a second invocation of
  \Cref{lem:localEst} or directly from \eqref{eq:gsEst}.

  {\em Ad (ii)}. This is another direct consequence of \Cref{lem:localEst}, which
  implies for equal times but distinct creation operators, with possibly distinct
  choices of scaling in the allowed region, that 
  \begin{align*}
    \norm{\Psi_{\T} - \Psi_{\T}'} &\leq
  C_1
  \sum_{j=1}^n \norm{\BB_{j\Oc\T}\Omega - \BB_{j\Oc\T}'\Omega}
  + C_2 \abs{\T}^{-\delta}, 
  \end{align*}
  where as before \( \delta:=  \kappa/4 - n\gamma\mu  > 0\). If \(\lim_\T \BB_{j\Oc\T}\Omega 
  = \lim_\T \BB_{j\Oc\T}'\Omega\), we obtain that the limits of \(\Psi_\tau\) and
  \(\Psi_\tau'\) coincide and that they are invariant under changes of scaling as claimed.
\end{proof}

\section{Fock structure of scattering states}
\label{sec:Fock}
Finally, we want to establish the Fock structure of scattering states, which
provides a simple formula for computing scalar products of any two scattering
states in terms of their single-particle components. An important consequence is
the non-vanishing of the limits defining the scattering states and it is the
essential ingredient to establish the extension of wave operators to the
full asymptotic Fock spaces (cf.\  \cite{Dy09}~App.~A).
With the clustering relation of creation-operators of \Cref{cor:cluster} at hand, 
the arguments leading to the Fock structure of scattering states are well-known
and we can not refrain from rephrasing them, e.g.\ from~\cite{Dy05}.
We will use the abbreviation \([n]:= \{1,2,\ldots,n\}
\subset\NaturalNum\) for finite subsets of natural numbers and 
\(\SymmetricGroup_n\) denotes the symmetric group of degree \(n\) in its
defining representation, i.e.\ acting on \([n]\).

We now consider two scattering state approximants (\(n, n' \in \NaturalNum_0\))
\begin{align*}
  \Psi_\T &: =  \BB_{1\Oc\T} \ldots \BB_{n\Oc\T} \Omega,
 & 
  \Psi'_\T &: =  \BB_{1\Oc\T}' \ldots \BB_{n'\Oc\T}' \Omega,
\end{align*}
such that \(\BB_{k\Oc\T}\) and \(\BB_{k\Oc\T}'\) have disjoint velocity supports within
each family.  Assuming finite Reeh-Schlieder degrees, the \term{outgoing} and
\term{incoming} \term{scattering states} \(\Psi^\pm := \lim_{\T\rightarrow \pm\infty} \Psi_\T\),
respectively, are well-defined by \Cref{thm:conv} for sufficiently small choices of scaling
\(\beta(\T)=\abs{\T}^{-\mu}\), \(\mu > 0\), and similarly for \(\Psi'^\pm :=
\lim_{\T\rightarrow \pm\infty} \Psi'_\T\).
We denote the corresponding single-particle states by
\( \Psi_{k}^{(1)} := \lim_{\T\rightarrow \infty} \BB_{k\Oc\T}\Omega \), (\(1\leq k \leq n\))
and
\( \Psi_{k'}'^{(1)} := \lim_{\T\rightarrow \infty} \BB_{k'\T}'\Omega \),
(\(1\leq k'\leq n'\)).

\begin{Thm}[Fock structure]
The scalar products of any two outgoing scattering states of the above form are given by%
\footnote{As usual, the right-hand side of \eqref{eq:fock} is consistently
  interpreted for \(n > n'\), yielding vanishing scalar products also in
this case (as a consequence of the vanishing Kronecker delta \(\delta_{nn'}\)).}
  \begin{align*} 
    \left\langle \Psi^+, \Psi'^+ \right\rangle
    &
    = \delta_{n\Oc n'} \sum_{\pi \in \SymmetricGroup_n} 
    \prod_{k=1}^{n} \left\langle\Psi_{k}^{(1)},
    \Psi_{\pi\raisebox{0em}[0em]{\tiny$(k)$}}'^{(1)}\right\rangle, 
    \label{eq:fock}\numberthis
  \end{align*}
  and similarly for incoming states.
\proof
  For simplicity we treat only the outgoing case \(\T \rightarrow +\infty\). 
  By continuity of the scalar product, the left-hand side of \eqref{eq:fock} can
  be written as the limit \(\T\rightarrow  \infty\) of
  \begin{align*}
    \left\langle \Psi_\T, \Psi'_\T \right\rangle 
      &= \left\langle  \BB_{1\Oc\T}\ldots \BB_{n\Oc\T} \Omega, 
        \BB_{1\Oc\T}'\ldots \BB_{n'\Oc\T}'\Omega\right\rangle,
    \label{eq:finiteT}\numberthis 
  \end{align*}
  where we can assume identical scaling \(\mu > 0\) for both sides by
  \Cref{thm:conv} (ii). Now we perform induction with respect to the number
  of particles~\(n'\) (assuming without restriction that \(n' \geq n\)).
  For each \(n'\) and \(n = 0\), statement \eqref{eq:fock} is equivalent to
  \(\norm{\Omega}=1\) for \(n = 0\)
  and \( \left\langle  \Psi'^+, \Omega \right\rangle = 0 \) for \(n' > 0\).
  The latter follows from  \cref{eq:finiteT} and the spectral support argument
  of \Cref{prop:elemProp}~\itref{it:vacAnnihil}.

  Assuming now that \eqref{eq:fock} holds for \(n-1\) particles, one can show by
  means of \Cref{lem:doubleComm} and \Cref{cor:cluster} that, up to terms
vanishing for \(\abs{\T} \rightarrow \infty\),  \eqref{eq:finiteT}  equals
  \begin{align*}
    \fl[5em]
    \sum_{k=1}^{n'}
    \left\langle \Omega, \BB_{n\Oc\T}^*\ldots \BB_{2\Oc\T}^* \BB_{1\Oc\T}' \ldots \BB_{k-1\NOc\T}' 
      \BB_{k+1\NOc\T}' \ldots \BB_{n'\Oc\T}' E_\Omega \BB_{1\Oc\T}^*  \BB_{k\Oc\T}'\Omega\right\rangle
   \\& \overset{\mathclap{\T \rightarrow \infty}} \longrightarrow \;\; 
     \sum_{k=1}^{n'} \Bigg(
     \bigg(\delta_{n-1, n'-1} 
       \!\!\!\!\sum_{\pi \in \SymmetricGroup_{n-1}(1,k)} \prod_{l=2}^{n}
       \left\langle\Psi_{l}^{(1)},\Psi'^{(1)}_{\pi\raisebox{0em}[0em]{\tiny$(l)$}}\right\rangle
     \bigg) \cdot 
     \left\langle \Psi_{1}^{(1)}, \Psi_{k}'^{(1)}\right\rangle \Bigg),
  \end{align*}
  where \(\SymmetricGroup_{n-1}(1,k)\) denotes the set of bijective
  maps~\(\pi\) between the two sets of numbers~\([n] \setminus \{1\}\) and
  \([n]\setminus \{k\}\) 
  and convergence is inferred from the induction assumption.
  Note that while \(\SymmetricGroup_{n-1}(1,k)\) is by itself not a group (its elements
  are maps between different sets and thus cannot be composed), it can
  nevertheless be identified with the subset of \(\pi \in \SymmetricGroup_{n}\)
  for which \(\pi(1) = k\). This implies that
  \begin{align*}
  \lim_{\T\rightarrow\infty}
    \left\langle \Psi_\T, \Psi'_\T \right\rangle
    &= \delta_{n\Oc n'}
  \sum_{k=1}^n \sum_{\substack{\pi \in \SymmetricGroup_n \\ \pi(1)=k}} \prod_{l=1}^{n}
  \left\langle\Psi_{l}^{(1)},\Psi'^{(1)}_{\pi\raisebox{0em}[0em]{\tiny$(l)$}}\right\rangle 
  =
  \delta_{n\Oc n'} \sum_{\pi \in \SymmetricGroup_n} \prod_{l=1}^{n}
  \left\langle\Psi_{l}^{(1)},\Psi_{\pi\raisebox{0em}[0em]{\tiny$(l)$}}'^{(1)}\right\rangle 
       .\qedhere
  \end{align*}
\end{Thm}

\section{Conclusions and outlook}
\label{sec:disc}

We have established the existence and Fock structure of scattering states
corresponding to 
single-particle states \(\Psi_1 \in E(H_m)\HilbertSpace\) with finite
Reeh-Schlieder degree. This assumption requires  the existence of a family
of local operators~\((\Aa_\beta)_{\beta > 0} \subset \Alg(\Reg)\) such that
\[
  \norm{\Aa_\beta\Omega - \Psi_1} \leq \beta,\qquad \norm{\Aa_\beta} \leq
  \beta^{-\gamma}.
      \label{eq:condFin} \tag{RS} 
\]

Beyond \eqref{eq:condFin} our method has no further dependence on the
concrete mechanism (e.g.\ additional ergodic averaging as in \cite{Dy05}) yielding a
limit of \(\Aa_\beta \Omega\) in the single-particle space. We have seen that
the Haag-Ruelle construction can be adapted, so that any finite degree \(\gamma
\) is feasible. Thus an arbitrarily strong polynomial growth of
\(\norm{\Aa_\beta}\) relative to the convergence of \(\Aa_\beta\Omega\) to the
single-particle vector \(\Psi_1\) can be handled. 

As mentioned in the introduction, Assumption~\eqref{eq:condFin} is readily 
verified in free field theory (cf.\ also \Cref{app:gff}). Its status in concrete
interacting models or within the general axiomatic framework is beyond the scope
of the present work and poses an interesting problem for future research.
We will briefly summarize our current understanding regarding the validity of
conditions of strengthened Reeh-Schlieder type and also give some additional
supporting arguments for our approach to the construction of scattering states.
We shall refrain from going into technical details, as we intend to provide them
elsewhere.

\begin{enumerate}[(a)]
 \item  \label{it:gap} 
    Quantitative improvements in the construction of scattering states regarding
    the strength of condition \eqref{eq:condFin} are possible.  Most notably in
    theories with lower mass gap one can show that already
    \((\Aa_\beta)_{\beta > 0} \subset \Alg(\Reg)\),
    \begin{align*}
      \norm{E(\Delta)(\Aa_\beta\Omega - \Psi_1)} \leq C_\Delta \beta,\qquad \ln \norm{\Aa_\beta} \leq
      \beta^{-\gamma},
      \label{eq:condFlat} \tag{RS\(^\flat\)}
    \end{align*}
    is sufficient for establishing scattering theory.
    Here \(\Delta \subset \RealNum^{\sPOne}\) is an arbitrary compact set, and
    \(C_\Delta > 0\) does not depend on \(\beta\).
    Intuitively, the stronger norm increase in \eqref{eq:condFlat} may be 
    compensated by the exponential space-like clustering in these models.

  \item 
    It was already pointed out that previous constructions of scattering states of
    embedded (massive) particles commonly need to assume additional regularity
    of the spectral measure near the particle masses. Here we briefly comment on
    the relation of such regularity assumptions to conditions of Reeh-Schlieder
    type. For spectral regularity according to Herbst, one requires there exist local
    operators \(A\in\Alg(\Reg)\) such that in addition to a nonvanishing
    single-particle
    component~\(E_mA\Omega\), one has for a suitable \(\epsilon > 0\) and all
    small enough \(\delta > 0\),
    \cite{Hrb71,Dy05}%
    \footnote{Weakened variants of \eqref{eq:herbst} have also been discussed
      recently, see e.g.\ 
    \cite{Hdg13, DH14}.}
    \begin{align*}
      \norm{E(H_m^\delta \setminus H_m) A \Omega} 
      \leq C \delta^{\epsilon}, \quad \text{where } H_m^\delta := \quad
      \bigcup_{\mathclap{\abs{\mu - m } < \delta}} \;H_\mu,
      \label{eq:herbst}
      \tag{H}
    \end{align*}
    and that the set of single particle vectors  obtained from
    such operators is dense in the single particle space \(E_m\HilbertSpace\).

    Starting from an operator \(A \in \Alg(\Reg)\) as in \eqref{eq:herbst},
    one can show by a very crude but general construction using differential
    operators that there exists a dense set of single particle states \(\Psi_1
    \in E_m \HilbertSpace\), which are generated by operators 
    satisfying \eqref{eq:condFlat}, with \(\gamma > 0\) inversely proportional
    to the Herbst constant \(\epsilon\) from~\eqref{eq:herbst}.
    Here we do not even need to invoke the Reeh-Schlieder property --- one may make use of
    the non-local nature of the energy-projection~\(E(\Delta)\) in
    condition~\eqref{eq:condFlat} to generate
    single-particle states (even if \(\Delta\) is larger than a subset of the mass
    hyperboloid).  Improving
    upon this result appears to require a more
    detailed quantitative understanding of the non-local correlations implied by
    the Reeh-Schlieder theorem, which may be model-dependent --- cf.\ also
    \Cref{app:gff}.

   \item \label{it:grow}
    We restricted our analysis to uniformly localized Reeh-Schlieder families solely for
    technical convenience.
    The present method may be refined to admit families \(\Aa_\beta \in
    \Alg(\DoubleCone_{R_\beta})\) similarly as in \eqref{eq:condFin}, but
    localized in double cones \(\DoubleCone_{R_\beta}\) of polynomially growing
    radii~\(R_\beta : = \beta^{-N}\) (for some \(N>0\)).

    A similar delocalization commonly enters in previous approaches
    via ergodic averaging prescriptions \cite{Hrb71,Dy05,Hdg13,DH14}.
    Due to the geometrical limitations discussed in \Cref{sec:commut},
    this delocalization appears to necessitate
    Herbst-type spectral conditions \cite{Hrb71} in these works. 
    Allowing a weakened localization \(\Aa_\beta \in
    \Alg(\DoubleCone_{R_\beta})\) might help to understand the relation of such
    spectral conditions to the Reeh-Schlieder
    condition~\eqref{eq:condFin}.
\end{enumerate}

\def\Mm{\mathcal M}
\def\EE{\mathcal E}
\def\Hh{\mathcal H}
\def\Ww{\mathcal W} 
A more concrete investigation of \eqref{eq:condFin} can be carried out using the
concept of polarization-free generators \cite{BBS01}. In this setting, we can
derive a wedge-local variant of the Reeh-Schlieder condition from
the domain condition 
\(\Omega \in \Dom(T^{1+\epsilon})\) for some~\(\epsilon > 0\), where \(T \geq
0\) denotes the self-adjoint part of the polar decomposition of a suitable
polarization-free generator~\(G = UT\). With this
input we can proceed as in free field theory and set \(A_\beta : = UT\Ee^{-\beta
T^{\epsilon}}\)
to obtain {\em wedge-local}\/ Reeh-Schlieder families of degree \(\gamma = \epsilon^{-1}\).
If a correponding variant of \Cref{conj:AHR} holds for oppositely localized
pairs of such wedge-local operators, as it is the case in purely massive
theories \cite{Fre85}, our results may be extended to yield a
construction of two-particle scattering states for embedded Wigner particles.

In this setting, it is problematic to imitate the Haag-Ruelle
construction by directly smearing polarization-free generators~\(G\) due to the
complicated structure of the domains~\(D(G)\).  It has been shown that even
ostensibly weak temperateness assumptions with
respect to the action of space-time translations on \(\Dom(G)\) imply triviality
of scattering in massive theories on Minkowski space with spatial dimension~\(s
> 1\) \cite{BBS01}. Therefore it is a subtle question whether the above domain
condition is compatible with non-trivial scattering.\footnote{%
  Preliminary computations suggest that \(\Omega \in \Dom(T^{1+\epsilon})\)
  could be fulfilled in certain \(1\!\!+\!\!1\)-dimensional
  integrable models with non-temperate polarization free generators
  \(G\) \cite{CT15} [Yoh Tanimoto, private communications]. A definite assessment requires
  the construction of a Borchers triple for these models, which has not yet been
  completed at the time of writing of this work.
} 

\appendix

\section{Notation and Conventions}
\def\sPOnex{4}
\def\sPOnexH{2}
\def\sPTwo{5}

For the Minkowski space-time metric we use the convention \(k \cdot x := k^0 x^0 - \vec k \cdot \vec x\)
for \(k, x \in \RealNum^{\sPOne}\).
Accordingly, the Fourier transform of a Schwartz functions \(f \in
\SchwartzSpace(\RealNum^{\sPOne})\) is defined by
  \[
    \hat f(k) : = \frac{1}{(2\pi)^2}
  \int \DInt[4] x \;\Ee^{\Ii k \cdot x} f(x).
\label{eq:ft}\numberthis
\]
The wave-packet \(\tilde f\) of a regular Klein-Gordon solution \(f\) (as
  defined in \Cref{sec:basic}),
is related to a corresponding partial (spatial) inverse transform of \(f_t(\vec x):= f(t, \vec
x)\) at \(t = 0\) by a factor~\((2\pi)^{\s/2}\).

The Fourier transform on the extended space \(x = (\vec x,s)\) and space-time    \(\uvec x =
     (x^0, \vec x,s)\) (see \Cref{app:gff})
is defined  for  \(\uvec f \in \SchwartzSpace(\RealNum^{\sPTwo})\)
     and \({\bf f} \in \SchwartzSpace(\RealNum^{\sPOnex})\) 
     by
     \begin{align*}
  {\uvec{\hat f}}(\omega, \vec k, \mu) 
  &: = \frac{1}{(2\pi)^{\sPTwo/2}}\;
    \int \DInt[\sPTwo]x \;\Ee^{\Ii \omega x^0 - \Ii \vec k \cdot \vec x - \Ii \mu s}
   \, \uvec f(x^0, \vec x, s),
    \\
  \hat{\bf  f}( \vec k, \mu)
  &: = \frac{1}{(2\pi)^{\sPOnexH}}\;
    \int \DInt[\sPOnex]x \;\Ee^{- \Ii \vec k \cdot \vec x - \Ii \mu s}
    \, {\bf f}( \vec x, s).
  \end{align*}

For \(x = (t, \vec x) \in \RealNum^{\sPOne}\) we write
\(
  A(x) := \alpha_x(A) : = U(x)AU(x)^*
\)
and similarly for \(\alpha_t(A)\) and
\(\alpha_{\vec x}(A)\).
By weak integration, these automorphisms of the global algebra
induce for given \(A\in\Alg\)  (regular) operator-valued
distributions 
\[
  A(f) := \int \DInt[\sPOne] x  \; f(x)  \alpha_x(A), \quad
  f\in\SchwartzSpace(\RealNum^\sPOne)
\]
and similar distributions \(A(g)\) are obtained for spatial smearing with \(g \in
\SchwartzSpace(\RealNum^\s)\). 

\section{Uniformly almost-local operator \tfamilies} \label{app:proofs}
An operator \(A \in \Alg\) is \term{almost-local} if there exists
for any \(r>0\) a double-cone localized operator \(A_r \in
\Alg(\DoubleCone_r)\),
such that for each \(N \in \NaturalNum\) with a suitable constant \(C_N\) we
have
\[
  \norm{A - A_r} \leq \frac{C_N}{(1+r)^N}. \label{def:almLoc} \numberthis
\]
For certain families \((A_\beta) \subset \Alg\) of almost-local operators, the
behaviour of corresponding constants \(C_{N,\beta}\) in \eqref{def:almLoc} with
respect to the parameter \(\beta > 0\) can be quantified in a simple manner.

\begin{Prop}
  \label{lem:unifAlmLoc} Let \(\Aa_\beta \in \Alg(\Reg)\) (\(\beta > 0\)) be an
  operator \tfamily\ localized in a fixed bounded region~\(\Reg \subset \RealNum^\s\) and
  let \(\chi \in \SchwartzSpace(\RealNum^{\sPOne})\). Then the \tfamily\ of
  almost-local operators \(\Bb_\beta := \Aa_\beta(\chi)\) is {\em uniformly
  almost-local} relative to \(\norm{\Aa_\beta}\) in the following sense:
  for each~\(\beta > 0\) there are 
  \(\Bb_{\beta, r} \in \Alg(\DoubleCone_r)\) (\(r > 0\)), such that
  for all \(N\in \NaturalNum\) 
  \[ 
   \exists C_N > 0
    \;\forall \beta > 0:
    \norm{\Bb_\beta - \Bb_{\beta,r}} \leq \frac{C_N\norm{\Aa_\beta}}{1+r^N}.
    \label{eq:unifAL}\numberthis
  \]
  Notably, the constants \(C_N\) are uniform in \(\beta\).  This also implies
  \begin{align*}
   \int \DInt[\s] x \norm{\left[\Bb_{\beta}, \Bb_{\beta}^*(\vec
   x)\right]} \leq  C_{\chi,\Reg} \norm{\Aa_{\beta}}^2.
 \label{eq:ALInt}\numberthis
  \end{align*} 
 \proof Let us assume for concreteness that \(\Aa_\beta \in
 \Alg(\DoubleCone_R)\) with the double-cone radius~\(R > 0\) fixed. 
 As \(\chi\in \SchwartzSpace(\RealNum^{\sPOne}) \),
 we obtain natural candidates for approximating local
 operators
  \[
    \Bb_{\beta, r} := 
    \int\limits_{\mathclap{\abs{x}_c < r - R}} \DInt[\sPOne] x \; \chi(x) \Aa_\beta(x)
    \in \Alg(\DoubleCone_r)
  \]
  (for \(r \leq R\) we simply set \(\Bb_{\beta,
  r} = 0\)). By the rapid decay of \(\chi\), we get for \(r > 2R\),
  \begin{align*}
    \norm{\Bb_\beta - \Bb_{\beta, r}} &  
    \leq \norm{\Aa_\beta}
    \cdot \int\limits_{\mathclap{\abs{x}_c \geq r - R}} \DInt[\sPOne] x \;
    \abs{\chi(x)}
    \leq \frac{C_N \norm{\Aa_\beta}}{1+(r-R)^N}
    \leq \frac{C_{N,R}' \norm{\Aa_\beta}}{1+r^N}.
  \end{align*}
  Together with the trivial estimate \(\norm{\Bb_\beta} \leq \norm{\Aa_\beta}
  \norm{\chi}_1\) for \(r \leq 2R\), this implies \eqref{eq:unifAL}.

  To obtain \eqref{eq:ALInt} we use an \(\abs{\vec x}\)-dependent local
  approximation \(\Bb_{\beta,r}\) under the integral: 
  choosing \(r = r(\vec x) : = \abs{\vec x}/2\) the commutator 
  \([\Bb_{\beta,r(\vec x)}, \Bb_{\beta,r(\vec x)}^*(\vec x)]\) will vanish by
  locality and thereby we have reduced the integrand to terms proportional to
  the approximation error. More explicitly we rewrite the left-hand side as
  \begin{align*}
   \int \DInt[\s] x \norm{\left[(\Bb_{\beta} - \Bb_{\beta,r(\vec x)})  + \Bb_{\beta,r(\vec x)} , 
   (\Bb_{\beta}^*(\vec x) - \Bb_{\beta,r(\vec x)}^*(\vec x))  +
   \Bb_{\beta,r(\vec x)}^*(\vec x) \right]}.
  \end{align*}
  After expanding the commutator (preserving the two differences in brackets)
  and utilizing subadditivity,
  \(\normm{[ \Bb_{\beta,r(\vec x)},  \Bb_{\beta,r(\vec x)}^*(\vec x) ]}\)
  vanishes for all \(\vec x\) by construction (due to locality). All remaining terms will
  contain at least one difference
  \(\Bb_\beta - \Bb_{\beta,r(\vec x)}\) or its translate.
  Using \eqref{eq:unifAL} we can now directly
  estimate the integral, 
  \[
    \norm{\left[\Bb_{\beta} - \Bb_{\beta,r(\vec x)} , \Bb_{\beta,r(\vec x)}^*(\vec x) \right]} 
    \leq 2 \norm{\Bb_{\beta} - \Bb_{\beta,r(\vec x)}}\norm{ \Bb_{\beta,r(\vec
      x)}^*(\vec x)}
      \leq \frac{2 C_N \norm{\Aa_\beta}^2}{1+r^N}.
  \]
  Taking \(N\) sufficiently large we obtain convergence of the integral and~\eqref{eq:ALInt}.  \qedhere \end{Prop}
  \section{Reeh-Schlieder Families in Generalized Free Models} \label{app:gff}
  \def\sPTwo{5}
  Let us briefly discuss the status of condition \eqref{eq:condFin} for
  noninteracting theories with embedded mass shell. Generalized free theories
  have proven useful to study Herbst-type spectral conditions \eqref{eq:herbst}
  (\cite{Dy05}, Sec.~4, see also \cite[Ch.\ 3.3, esp.\ p.\ 264 ff.]{Ho90} for a
  general review), and we think that the following considerations might also
  give some hints concerning strengthened Reeh-Schlieder properties in
  interacting theories\footnote{%
      Due to vacuum polarization \(\phi(f)\Omega\) cannot have sharp mass for
      interacting theories, i.e.\ there is some spectral background \(E_m^\perp \phi(f)\Omega \not = 0\).
      Generalized free fields simulate this in a simplistic way via~\eqref{eq:defRho}.}.
    The generalized free field \(\phi(f)\), \(f \in
    \SchwartzSpace(\RealNum^{\sPOne})\),
    may be interpreted as a certain superposition of ordinary free fields
    \(\phi_\mu(f)\) of mass \(\mu \geq 0\) with weight
    measure~\(\DInt\rho(\mu)\) describing the mass spectrum of the theory.
    For our purposes, \(\rho\) should consist of a delta measure at the desired
    particle mass~\(m \geq 0\) and some continuous background spectrum. 
    We will take
    \[
      \rho := \delta_m + \rho_{\text{cont}}, \quad
      \rho_{\text{cont}}(\Delta) := 
      \int\limits_{\mathclap{\Delta \cap [0,m+1]}} \DInt \mu\;\frac{1}{\abs{\mu - m}^{1-\epsilon}} + \alpha \lambda(\Delta),
      \label{eq:defRho}\numberthis
    \]
    {for Borel sets} \(\Delta \subset [0,\infty)\),
    where \(\lambda\) denotes Lebesgue measure.
     The parameter \(\epsilon > 0\) controls the regularity in
     the vicinity of the particle mass, i.e.\ regarding the Herbst condition
     \eqref{eq:herbst}. Additionally, the support properties of \(\rho\), governed by
     \(\alpha \in \{0, 1\}\), are of (perhaps unexpected) relevance for the
     Reeh-Schlieder problem.

    On the bosonic Fock space \(\FockSpace_\rho :=
    \Gamma(\HilbertSpace_{1,\rho})\) over the single-particle space~\(\HilbertSpace_{1,\rho}
    := \LSpace^2(\RealNum^\s) \otimes \LSpace^2([0,\infty), \DInt \rho)\) we
    obtain a Wightman field in terms of the Segal operators
    \(\Phi_S(\psi) := (a^*(\psi) + a(\psi))/\sqrt{2}\), \(\psi \in
    \HilbertSpace_{1,\rho}\), 
       for real-valued test functions~\(f \in
    \SchwartzSpace_\RealNum(\RealNum^{\sPOne})\) by
    \[
      \phi(f) = \Phi_S(\omega^{-1/2} \hat f_+),
      \label{eq:defPhi}\numberthis
    \]
    where the argument contains the restriction
    \(\hat f_+ (\vec p, \mu) :=  \hat f(\omega_\mu(\vec p), \vec p)\),
    \(\omega_\mu(\vec p) := \sqrt{\vec p^2 + \mu^2}\), and
    \(\omega\) denotes the corresponding (unbounded) multiplication
    operator on \(\HilbertSpace_{1,\rho}\). The representation of translation
    group is generated by the second quantization of the multiplication
    operators~\((\omega, \vec p)\),
    and setting \(W(f) : = \Ee^{\Ii \phi(f)}\), we obtain a corresponding Haag-Kastler net 
    for bounded open regions \(\Reg \subset \RealNum^{\sPOne}\) by
    \[
      \Alg(\Reg) := \{ W(f): \;
	f \in \SchwartzSpace_\RealNum(\RealNum^{\sPOne}), 
	  \; \support f \subset  \Reg\}''.
          \label{eq:gffnet}\numberthis
    \]

    It will be convenient to adopt Landau's formulation \cite{La74}, as it gives
    a simple reinterpretation of \(\Alg(\Reg)\) in terms of time-zero fields.
    For Schwartz test functions \(\uvec f(\uvec x)\),
    \(\uvec x = (x^0 , \vec x, s)\), from here on assumed to be symmetric in
    \(s\), where~\(s\) may be interpreted as new auxiliary
    space-like\footnote{However the field \(\uvec
      \phi(\uvec f)\) should not be expected to be local  in the direction of
    \(s\).} variable conjugate to the mass \(\mu\), set
    \(
      \uvec\phi(\uvec f) := \Phi_S(\omega^{-1/2} \uvec{\hat f}_+) 
    \),
    \(\uvec{\hat f}_+(\vec p, \mu) :=  \uvec{\hat f}(\omega_\mu(\vec p),
    \vec p, \mu)\). 
    Analogously to~\eqref{eq:gffnet}, we obtain an extended net
    \(
      \uvec\Alg(\uvec \Reg)
    \)  on~\( \RealNum^{\sPTwo}\).

    It is easily seen that extended field~\(\uvec \phi(\uvec
f)\) and its time derivative \(\uvec \phi_t(\uvec f) := - \uvec \phi(\partial_t
\uvec f)\) admit well-defined restrictions to time-zero fields 
    \begin{align*}
      \uvec\phi_0(\bf f) &= 
      \Phi_S(\omega^{-1/2} \, {\hat{\bf f}})
      ,
      &
      \uvec\pi_0(\bf f) &= 
      \Phi_S(\Ii\omega^{1/2} \, {\hat{\bf f}})
      \label{eq:extt0}\numberthis
    \end{align*}
    for test functions \({\bf f} \in \SchwartzSpace(\RealNum^{\sPOne},
    \RealNum)\) defined on the extended \((\vec x, s)\)-space. 
 In terms of  corresponding extended double cones \( \uvec \DoubleCone_R := \{(t, \vec x,
 s) \in \RealNum^{\s+2}: \abs{t} + \sqrt{\vec x^2 + s^2} < R\} \), (\(R > 0\)),
 Landau gave the following characterization of the net \eqref{eq:gffnet}.
 \begin{Thm}\label{thm:gfft0} \cite{La74}.
    \(
     \Alg(\DoubleCone_R) = \uvec \Alg(\uvec \DoubleCone_R) 
    \). Furthermore, these algebras are generated by bounded functions of the
    time-zero fields  \eqref{eq:extt0} with test functions \({\bf f} \in
    \SchwartzSpace(\RealNum^{\sPOne}, \RealNum)\) supported in the
  ball~\(\Ball_R = \uvec \DoubleCone_R\big|_{t=0}\).
\end{Thm}
 
  \begin{Prop}\cite{La74}. \label{prop:cptRho}
   If the defining measure~\(\rho\) of the generalized free field is
   exponentially decreasing, then \(
     \Alg(\DoubleCone_R) = \uvec \Alg(\DoubleCone_R \times \RealNum) 
    \).
  \end{Prop}

  For choosing \(\alpha = 0\) in \eqref{eq:defRho}, we may  conclude that the
  strengthened Reeh-Schlieder property holds for the net \(\Alg\): 
  take a family of test functions \(\ff_\beta \in
    \ContinuousFuncs^\infty_c(\RealNum^{\sPOne})\), such that \(\{0\} \times \support \ff_\beta \subset \Reg
  \times \RealNum\) and with Fourier transforms
  converging sufficiently rapidly to a smooth limit supported on the sharp-mass
  subset~\(\RealNum^\s\times\{m\}\).
  By \Cref{prop:cptRho} we can make such a choice which is compatible with
  bounded functions of \(\phi_\beta := \uvec \phi_0(\ff_\beta)\),  such as
  \(A_\beta := \phi_\beta \Ee^{- \beta \abss{\phi_\beta}^N}\), being contained
  in the local algebra \(\Alg(\Reg)\), thus confirming the validity of~\eqref{eq:condFin}.
  Regarding \eqref{eq:condFin}
  we may summarize:
  \begin{Prop}
    For generalized free field models defined by \eqref{eq:defRho} with \(\alpha = 0\), 
    there exists a dense set of sharp-mass single-particle states generated by
    Reeh-Schlieder families of arbitrarily small degree \(\gamma > 0\)
    independently of the choice of \(\epsilon\) in \eqref{eq:defRho}.
  \end{Prop}
   A fortiori, a continuity argument then shows that the sharp-mass free field net
  \(\Alg_m(\Reg)\) is a subnet of \(\Alg(\Reg)\).  To obtain a non-trivial
  example we should thus choose \(\alpha = 1\). 
  We conclude with a short consideration of this difficult case, for which the
  assumptions of \Cref{prop:cptRho} are violated.
  
  Given a
   bounded double-cone region \(\DoubleCone_R\) and a single-particle 
   vector \(\Psi_1 \in \HilbertSpace_{1,\rho}\) (say \(\Psi_1 = \uvec\phi_0(\ff)\Omega\),
    with \(\ff \in \SchwartzSpace(\RealNum^{\sPOnex})\) supported in a very large
  region) we would like to find a family
   of smeared field operators \(\phi_\beta\) localized in \(\DoubleCone_R\),
   such that
   \( \norm{\phi_\beta\Omega - \Psi_1} \leq \beta \). For this purpose it will
   be convenient to introduce the following closed single-particle subspaces
   (\(\ff \in \SchwartzSpace(\RealNum^{\sPOnex})\)) in the setting of
   \Cref{thm:gfft0},
\[
    \HilbertSpace_{\uvec \phi_0,\Ball_R} := 
    \closure{
    \{ \phi_0(\ff)\Omega,\; \support \ff \subset \Ball_R\}},
        \;
          \HilbertSpace_{\uvec \pi_0,\Ball_R} := 
          \closure{
            \{ \pi_0(\ff)\Omega,\; \support \ff \subset
          \Ball_R\}}.
              \label{eq:defH1}\numberthis
            \]

        The orthogonal projections \(P_\phi\), \(P_\pi\) corresponding to
        \eqref{eq:defH1} may be used to iteratively define approximations of
        \(\Psi_1\) by vectors from \eqref{eq:defH1} or equivalently, generated
        by \(\DoubleCone_R\)-localized operators. Underlining 
        error terms after each half-step we begin with
        \[
          \Psi_1 = P_\phi \Psi_1 + \underline{(1-P_\phi)\Psi_1} = P_\phi \Psi_1
          + P_\pi P_\phi^\perp \Psi_1 + \underline{P_\pi^\perp
          P_\phi^\perp\Psi_1} = \ldots
        \]
        Similarly, after \(N\) iterations the remaining error is given by
        \(\normm{(P_\pi^\perp P_\phi^\perp)^N\Psi_1}\). By the von Neumann
        alternating projection theorem \cite[Thm.~13.7]{VN50}, \((P_\pi^\perp
        P_\phi^\perp)^N\) in fact converges strongly  to the orthogonal
        projection onto the intersection
        \(
          \HilbertSpace_{\uvec \phi_0,\Ball_R}^\perp \cap
          \HilbertSpace_{\uvec
              \pi_0,\Ball_R}^\perp = (
              \HilbertSpace_{\uvec \phi_0,\Ball_R} + \HilbertSpace_{\uvec
          \pi_0,\Ball_R})^\perp \). The latter is trivial by the Reeh-Schlieder theorem,
          implying convergence of our iterative procedure.
        An upper bound on the degree of sharp-mass Reeh-Schlieder
        families along the lines of \eqref{eq:condFin} or \eqref{eq:condFlat}
        may be inferred from the speed of convergence
        \(\normm{(P_\pi^\perp P_\phi^\perp)^N\Psi_1} \to
        0\), \(\Psi_1 \in E_m\HilbertSpace_{1,\rho}\) or equivalent geometrical
        information regarding the situation of \(\Psi_1\) in relation to the spaces \eqref{eq:defH1}. 
        This is presently still under investigation.

{
\newcommand{\ifprefchar}{\ifpunctmark{'}}
\sloppy
\printbibliography
}

\end{document}